\documentclass{aa.hack}
\usepackage{txfonts}
\usepackage{graphicx}
\usepackage{natbib}
\bibpunct{(}{)}{;}{a}{}{,}

\begin{document}

\title{Weak lensing mass reconstructions of the ESO Distant Cluster Survey \thanks{Based on
observations obtained at the ESO Very Large Telescope (VLT) as part of the Large Program 
166.A-0162 (the ESO Distant Cluster Survey).}}

\author{D. Clowe \inst{1,} \inst{2} \and P. Schneider \inst{1} \and A. Arag\'on-Salamanca \inst{3} 
\and M. Bremer \inst{4} \and G. De Lucia \inst{5} \and C. Halliday \inst{6,} \inst{7} \and 
P. Jablonka \inst{8} \and B. Milvang--Jensen \inst{9} \and R. Pell{\' o} \inst{10} \and 
B. Poggianti \inst{6} \and G. Rudnick \inst{5} \and 
R. Saglia \inst{9} \and L. Simard \inst{11} \and S. White \inst{5} \and D. Zaritsky \inst{2}}
\institute{
Institut f\"ur Astrophysik und Extraterrestrische Forschung der Universit\"at Bonn, Auf dem H\"ugel 71, 53121 Bonn, Germany \and
Steward Observatory, University of Arizona, 933 North Cherry Avenue, Tucson, AZ 85721 \and
School of Physics and Astronomy, University of Nottingham, NG7 2RD, UK \and
Department of Physics, Bristol University, H H Wills Physics Laboratory, Tyndall Avenue, Bristol, BS8 1TL, UK \and
Max--Planck--Institut f\"ur Astrophysik, Karl--Schwarschild--Str. 1, Postfach 1317, D-85741 Garching, Germany \and
Osservatorio Astronomico, vicolo dell'Osservatorio 5, 35122 Padova, Italy \and
Institut f{\"u}r Astrophysik, Friedrich-Hund-Platz 1, 37077 G{\"o}ttingen, Germany \and
GEPI, CNRS-RMS8111, Observatoire de Paris, section de Meudon, 5 Place Jules Janssen, F-92195 Meudon Cedex, France \and
Max--Planck--Institut f\"ur extraterrestrische Physik, Giessenbachstrasse, D-85748 Garching, Germany \and
Laboratoire de'Astrophysique, UMR 5572, Observatoire Midi-Pyrenees, 14 Avenue E. Belin, 31400 Toulouse, France \and
Herzberg Institute of Astrophysics, National Research Council of Canada, Victoria, BC V93 2E7, Canada}

\offprints{D.~Clowe,\email{dclowe@as.arizona.edu}}

\date{Received 01 01 3000 / Accepted 01 01 3000}

\abstract{
We present weak lensing mass reconstructions for the 20 high-redshift clusters in
the ESO Distant Cluster Survey.  The weak lensing analysis was performed on
deep, 3-color optical images taken with VLT/FORS2, using a composite galaxy
catalog with separate shape estimators measured in each passband.  We find that
the EDisCS sample is composed primarily of clusters that are less massive than those
in current X-ray selected samples at similar redshifts, but that all of the fields are
likely to contain massive clusters rather than superpositions of low mass groups.
We find that 7 of the 20 fields have additional massive structures which are
not associated with the clusters and which can affect the weak lensing mass
determination.  We compare the mass measurements of the remaining 13 clusters with
luminosity measurements from cluster galaxies selected using photometric redshifts
and find evidence of a dependence of the cluster mass-to-light ratio with redshift.  
Finally we determine
the noise level in the shear measurements for the fields as a function of
exposure time and seeing and demonstrate that future ground-based surveys
which plan to perform deep optical imaging for use in weak lensing measurements
must achieve point-spread functions smaller than a median of $0\farcs 6$ FWHM.
}

\titlerunning{Weak Lensing Reconstructions of EDisCS}
\authorrunning{Clowe et al.}

\maketitle

\section{Introduction}
The discovery of massive
($>10^{15} \mathrm{M}_\odot$), high-redshift ($z>0.6$) clusters in serendipitous X-ray
surveys \citep[eg the Einstein Medium Sensitivity Survey (EMSS),][]{GI90.1} resulted in constraints 
on the evolution of structure in the
Universe that excluded an Einstein-de Sitter model and favored a low mass-density
model \citep{OU92.1,EK96.1}.  Further constraints on cosmological models have come from
measuring the evolution, or lack there of, of various traits of the clusters, such as the 
X-ray gas temperature and gas mass fraction \citep{EV02.1,HE00.1}.  

These studies, however, have faced two potentially serious problems.  The first is that
the current X-ray surveys are only able to detect the most massive clusters at
high-redshift, which often results in a mismatch in the mass range between these high-redshift
clusters and the lower redshift samples to which they are compared.  The ESO Distant
Cluster Survey (EDisCS) is a sample of 20 fields likely to contain high-redshift clusters,
chosen from the optically selected Las Campanas Distant Cluster Survey \citep{GO01.1},
which detected clusters from smoothed overdensities in the sky level of shallow optical
exposures.  The LCDCS
covered roughly the same sky area as the deepest pointings in the EMSS but found an order
of magnitude more high-redshift clusters. As a result, the clusters in EDisCS should, on average,
be of lower mass than the X-ray selected samples, and thus more directly comparable to
lower redshift samples.

The second problem is that the high-redshift clusters are being observed at a time when structure
formation is much more dynamic than it is today,  so it is uncertain how well properties such as the X-ray gas
temperature and cluster galaxy velocity dispersion, which depend upon the cluster
being in virial equilibrium, relate to the mass of the cluster.  Thus, to check for
evolution in such cluster properties, one needs a measurement of the mass of the
clusters using a method that does not require the cluster to be in a particular dynamical
state.  Weak gravitational lensing, in which one measures the distortion in the shapes of
background galaxies by the gravitational potential of the cluster, provides a means to do this.

In this paper we use the deep optical VLT/FORS2 imaging from EDisCS to measure the gravitational
shear field produced by the clusters and so derive mass estimates for them.
In Section 2 we discuss the optical image reduction and object detection.  In Section 3 we
discuss the weak lensing analysis and present measurements of the mass and optical
luminosity of the clusters.  Results for the intermediate redshift ($z\sim 0.5$) clusters are 
discussed in Section 4, and those for the higher redshift sample in Section 5.  Discussion
of the results and potential systematic errors in the sample are given in Section 6.  Throughout
this paper we assume a $\Omega_\mathrm{m} = 0.3, \Lambda \ = 0.7, H_0 = 70 \mathrm{km/s/Mpc}$
cosmology and give all errors as $1\sigma$ unless otherwise stated.

\section{Data Reduction}
\subsection{Image Reduction and Object Detection}
We use deep VLT/FORS2 images of the EDisCS clusters.  Details of image acquisition and
reduction process are described in \citet{WH04.1}.  A summary of the exposure times
and PSF sizes for each of the three passbands for all twenty cluster fields is given
in Table \ref{exptable}.  For the weak lensing analysis,
we make one alteration to the image reduction process compared to that used for the
photometric analysis, which is to perform a smoothed sky subtraction on the individual
input images before co-addition.  This is done by using the IMCAT 
(http://www.ifa.hawaii.edu/$\sim$kaiser/imcat) {\emph findpeaks} peak--finder inverted
to find local minima on the images after being smoothed with a $1\arcsec$ Gaussian.  
Two images of the minima are then created, one with the minima
values in the pixel locations where they are found and one in which
the pixels with minima are set to 1, and both are smoothed with a $6\farcs8$ Gaussian.  The smoothed
minima image is then divided by the smoothed weight image to produce an image of the sky,
which is subtracted from the original image.  This process removes all small scale
fluctuations in the sky from the input images along with extended stellar (and larger
galaxy) halos and any intra-cluster light.  This step was necessary to minimize
the error in determining the second moments of the surface brightness of the galaxies, 
for which we assume the sky background can be modeled with a linear slope across the galaxy.
An ultra-deep image was also created by
co-adding the input images from all three passbands, weighted by the inverse of the
square of the sky-noise in each image.

\begin{table*}[t]
\caption{Exposure times and PSF FWHM}
\begin{tabular}{lcccccc}
\multicolumn{7}{c}{$z\sim 0.5$ clusters} \\ \hline
cluster & \multicolumn{2}{c}{$I$} & \multicolumn{2}{c}{$V$} & \multicolumn{2}{c}{$B$} \\
name & $t_\mathrm{exp}$ (m) & FWHM (\arcsec ) & $t_\mathrm{exp}$ (m) & FWHM (\arcsec ) & 
$t_\mathrm{exp}$ (m) & FWHM (\arcsec ) \\ \hline
CLJ1018.8-1211 & 60 & 0.77 & 60 & 0.80 & 45 & 0.80 \\
CLJ1059.2-1253 & 60 & 0.75 & 60 & 0.85 & 45 & 0.80 \\
CLJ1119.3-1129 & 45 & 0.58 & 45 & 0.52 & 40 & 0.58 \\
CLJ1202.7-1224 & 45 & 0.64 & 45 & 0.75 & 45 & 0.70 \\
CLJ1232.5-1250 & 45 & 0.52 & 50 & 0.63 & 45 & 0.56 \\
CLJ1238.5-1144 & 45 & 0.54 & 45 & 0.63 & 45 & 0.61 \\
CLJ1301.7-1139 & 45 & 0.56 & 45 & 0.64 & 45 & 0.58 \\
CLJ1353.0-1137 & 45 & 0.58 & 45 & 0.68 & 45 & 0.60 \\
CLJ1411.1-1148 & 45 & 0.48 & 45 & 0.60 & 45 & 0.59 \\
CLJ1420.3-1236 & 45 & 0.74 & 45 & 0.80 & 45 & 0.62 \\
\hline \hline
\multicolumn{7}{c}{$z>0.6$ clusters} \\ \hline
cluster & \multicolumn{2}{c}{$I$} & \multicolumn{2}{c}{$R$} & \multicolumn{2}{c}{$V$} \\
name & $t_\mathrm{exp}$ (m) & FWHM (\arcsec ) & $t_\mathrm{exp}$ (m) & FWHM (\arcsec ) & 
$t_\mathrm{exp}$ (m) & FWHM (\arcsec ) \\ \hline
CLJ1037.9-1243 & 120 & 0.56 & 130 & 0.54 & 120 & 0.55 \\
CLJ1040.7-1155 & 115 & 0.62 & 120 & 0.72 & 120 & 0.62 \\
CLJ1054.4-1146 & 115 & 0.72 & 150 & 0.78 & 120 & 0.67 \\
CLJ1054.7-1245 & 115 & 0.50 & 110 & 0.77 & 120 & 0.79 \\
CLJ1103.7-1245 & 115 & 0.64 & 120 & 0.75 & 130 & 0.83 \\
CLJ1122.9-1136 & 115 & 0.65 & 115 & 0.70 & 120 & 0.71 \\
CLJ1138.2-1133 & 125 & 0.60 & 110 & 0.68 & 120 & 0.58 \\
CLJ1216.8-1201 & 115 & 0.60 & 110 & 0.72 & 130 & 0.68 \\
CLJ1227.9-1138 & 145 & 0.74 & 130 & 0.83 & 155 & 0.73 \\
CLJ1354.2-1230 & 115 & 0.66 & 120 & 0.70 & 125 & 0.70 \\
\hline
\end{tabular}
\label{exptable}
\end{table*}

We detected objects in the ultra-deep images using SExtractor \citep{BE96.1} set to find
groups of three contiguous pixels that are $1\sigma$ above the sky-level.
SExtractor was then used in two-image mode, with the ultra-deep image as the detection
image, to measure the magnitudes, in both a $1\arcsec$ radius aperture and to an isophotal
limit of $1\sigma$ in the ultra-deep image, for the objects in all three passbands.  
These produced catalogs with $\sim 110$ objects/sq.~arcmin for the intermediate redshift
clusters and $\sim 200$ objects/sq.~arcmin for the high redshift clusters, with the
majority of the objects being noise peaks.  We determined the significance and size of each 
object in each passband by convolving the images with a series of Mexican-hat
filters and determining the smoothing radius, $r_\mathrm{g}$, at which the filtered objects
achieved maximum significance, $\nu$.  The catalogs were then cut by removing all of the
objects that are measured as having a smaller smoothing radius than the stars in all of
the passbands, those which did not reach a significance of $10\sigma$ in at least one
passband, and those which had a bad or saturated pixel or image border within an aperture of 
radius $3 r_\mathrm{g}$.
For each surviving object, typically $\sim 40\%$ of the detected objects, we measured a local
sky level, flux within an aperture of radius $3 r_\mathrm{g}$, and a $50\%$
encircled light radius, $r_\mathrm{h}$, in each passband.  
Stars were separated from galaxies by selection in $r_\mathrm{h}$, with only objects
selected as stars in all three passbands being designated stars, and objects selected
as stars in only one or two passbands having their significances set to 0 for those passbands.

The next step was to measure the reduced shear $\vec{g}$, corrected for PSF smearing, for each
object in each passband separately.  The objects had their zeroth, second, and fourth moments, 
weighted by a Gaussian with $\sigma = r_\mathrm{g}$, measured and used to construct 
ellipticities, $\vec{e}$, and shear and smear tensors as per the KSB technique \citep{KA95.4}. 
{ Stellar ellipticities, shear, and smear tensors were measured over the full range of
weight function sizes used in the galaxy measurements, and in all subsequent steps the stellar
ellipticities and tensors used in the KSB corrections are those which have the same weight
function size as the galaxy being corrected.}
We determined the PSF ellipticity field by fitting the stellar ellipticities with a bi-cubic 
polynomial, which was then scaled by the ratio of galaxy to stellar smear tensors and 
subtracted from the galaxy ellipticities \citep{KA95.4}.  A shear susceptibility tensor, 
$\tens{P}^\gamma$ \citep{LU97.1}, was created for each galaxy.  Because $\tens{P}^\gamma$ is
extremely noisy, instead of calculating $\vec{g}_i = {\tens{P}^{\gamma}_{ij}}^{-1} \vec{e}_j$
for each object from its measured $\tens{P}^{\gamma}$ one needs to average the 
$\tens{P}^{\gamma}$s of galaxies with similar light profiles.  Since one expects that
the shear susceptibility tensor depends mainly on the size and shape of the object,
we fit each component of $\tens{P}^{\gamma}_{ij}$ as a bi-pentic polynomial of $r_\mathrm{g}$ 
and $e_j$ and use the values of the fitted components.  The final reduced shear
measurement for each galaxy was then created from a weighted mean of the reduced shear in each
passband, weighting by the square of $\nu$ for each passband with a minimum $\nu>10$ required
in a passband for inclusion in the mean.  

We created a background galaxy catalog from the full galaxy catalog by applying color
cuts to eliminate probable cluster or foreground galaxies, a magnitude cut of
$V>23$, and a requirement of having $\nu > 10$ in at least one passband.  The color cuts were 
chosen to remove all cluster galaxies and most foreground galaxies while preserving
higher redshift ($z>0.9$) galaxies.  The number densities of the background galaxy
catalogs are given in Tables \ref{lztable} and \ref{hztable}.  A discussion of the number
densities of the background galaxies detected in each passband along with their contribution
to the combined catalog is given in Appendix A.  { Simulations using the PSFs in our images show
that this method tends to underestimate the true shear in the images by $\sim 4\%$ on average, 
although the exact amount is highly dependent upon the models used for the background galaxies.  
Because the spread in correction factors based on the background galaxy models used is comparable
to the mean correction factor in the simulations and the error in the weak lensing masses for the
clusters in this sample is typically an order of magnitude larger than this correction factor, 
we have chosen not to apply a correction factor to our measured shear values.  We instead present
the results with the caveat that we are likely underestimating the true masses of the clusters
by a few percent.
}

\subsection{Weak Lensing Analysis}

The goal of weak lensing analysis of a cluster field is to determine the convergence, $\kappa$,
across the field, which is related to the surface mass density, $\Sigma$, via
\begin{equation}
\kappa = {\Sigma \over \Sigma_{\mathrm{crit}}}.
\end{equation}
$\Sigma_{\mathrm{crit}}$ is a scaling factor:
\begin{equation}
\Sigma _{\mathrm{crit}}^{-1} = {4 \pi  G \over c^2}{ D_{\mathrm{ol}}
D_{\mathrm{ls}} \over D_{\mathrm{os}}}
\end{equation}
where $D_{\mathrm{os}}$ is the angular distance from the observer to the source (background)
galaxy, $D_{\mathrm{ol}}$ is the angular distance from the observer to the lens (cluster),
and $D_{\mathrm{ls}}$ is the angular distance from the lens to the source galaxy.
With knowledge of the redshift of the cluster, redshift distribution of the background
galaxies used to derive $\kappa$, and a chosen cosmology, one can convert a $\kappa$
distribution into a surface mass density distribution for the cluster.
What is measured from the background galaxies, however, is the reduced shear
\begin{equation}
\vec{g} = {\vec{\gamma} \over 1 - \kappa}
\end{equation}
which is a function of both the convergence and the gravitational shear, $\vec{\gamma}$.
To convert the measured reduced shear field around the clusters to a $\kappa$ distribution, we
use three different methods.

\begin{figure}
\sidecaption
\includegraphics{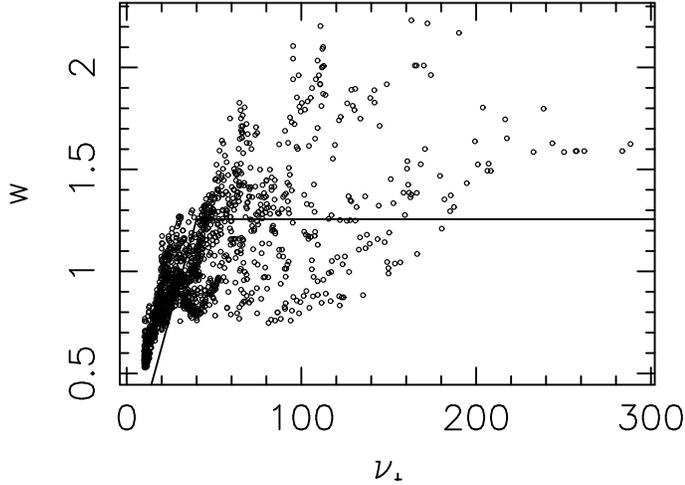}
\caption{Shown above are a plot of weights assigned to background galaxies in the CLJ1232$-$1250
cluster.  The plotted points show a weight calculated from the inverse rms of $g$ for the 50
nearest neighbors to each galaxy in a magnitude--size plane while the line shows the weight
using $\min (\nu _\mathrm{t}, 40)$, with both distributions normalized to give an average weight
of 1.}  
\label{weightfig}
\end{figure}

The first method produces a two-dimensional map of $\kappa$ using a finite-field inversion
with Neumann boundary conditions \citep{SE96.3}.  This technique utilizes
the fact that both $\kappa$ and $\vec{\gamma}$ are combinations of second derivatives of
the surface potential, and therefore 
\begin{equation}
\nabla \ln (1 - \kappa) = {1\over 1 - g_1^2 - g_2^2} \left( {\begin{array}{cc}
1 + g_1 & g_2 \\ g_2 & 1 - g_1 \end{array}} \right) \left ( {\begin{array}{c}
g_{1,1} + g_{2,2} \\ g_{2,1} - g_{1,2} \end{array}} \right)
\end{equation}
(Kaiser 1995), and solves for $\ln (1 - \kappa)$ within a field except for an unknown
additive constant.  The technique also requires a continuous field for both $\vec{g}$
and its first derivative, which we created by using a regularly spaced grid for the
shear field, with the shear value at each position being the weighted mean of the
shear values of the surrounding galaxies.  The weighting function used was
\begin{equation}
w = \exp\left({d^2\over 2 d_0^2}\right) \times \min (\nu _\mathrm{t}, 40)
\end{equation}
where $d$ is the distance between the grid point and the galaxy, $d_0$ is the chosen smoothing
length ($25\farcs 6$ for the reconstructions shown in Figs.~2--21), and $\nu_\mathrm{t}$
is the sum in quadrature of the significances in each passband.  The $25\farcs 6$ smoothing
length for the reconstructions shown in the figures was chosen as a compromise between wanting 
to suppress small-scale noise peaks in the mass reconstructions while also preserving the shape
of the mass distribution of the clusters in the inner few hundred kpc.  

The capping of 
$\nu_\mathrm{t}$ in the weight function was performed to prevent the shear field
from being dominated by a handful of small, bright, and presumably fairly low redshift 
galaxies, and the chosen value (40) was determined by finding where the root mean square
of the shear measurements began to increase with decreasing significance.  { This is
illustrated in Fig.~\ref{weightfig} where we have plotted our chosen weight formula, 
$\min (\nu _\mathrm{t}, 40)$, and the inverse of the rms dispersion of $g$ for each galaxy,
calculated from its 50 nearest neighbors in magnitude and size, versus $\nu _\mathrm{t}$
for CLJ1232.5$-$1250.
At small values of $\nu _\mathrm{t}$, the weight from the inverse rms of $g$ for the
majority of galaxies decreases linearly with $\nu _\mathrm{t}$, which is due to the increase in the
measurement noise of the second moments with decreasing galaxy brightness.  The large scatter
in the inverse rms of $g$ at large $\nu _\mathrm{t}$ is due partly to sampling noise from using
only 50 galaxies to measure the rms and partly to the most strongly lensed galaxies, which have
higher mean values of $g$ due to being more strongly lensed, tending to occupy a different part
of the magnitude, size plane compared to galaxies which are more weakly lensed.  This latter effect
also means that if one were to use the inverse rms of $g$ as a weight, the most strongly lensed
galaxies will tend to be given low weights, and thereby biasing the lensing measurements toward
low cluster mass.
The use of our chosen weight function increased the significance of the weak lensing detections
by an average of $\sim 30\%$ over both the unweighted measurement and a measurement using
the inverse rms of $g$ as the weight.  We note that our chosen weighting scheme may bias the
weak lensing measurements if there is a relation between galaxy ellipticity and $\nu _\mathrm{t}$.
Simulations, however, indicate that any such bias is smaller than the $1\%$ level.
}

We converted the $\ln (1 - \kappa)$ solutions into the $\kappa$ distributions shown in
Figs.~2--21 by assuming that $\kappa$ is zero at the edge of the fields.  The true solution
is
\begin{equation}
\kappa_{\mathrm{real}}(\vec{\theta}) = (1 - \lambda) \kappa_{\mathrm{obs}}(\vec{\theta}) + \lambda
\end{equation}
over the field $\vec{\theta}$ where $\lambda$ is the mean $\kappa$ at the edge of the field.
Because of the small size of the fields, $\kappa$ will still be above the cosmic mean at the
edges of the field, and our surface density maps will be too sharply peaked compared to the real
mass distribution.  However, this uncertainty should be only a few percent effect and well 
within the noise
level from the intrinsic ellipticity distribution of the background galaxies.

The second method produces a best-fit mass model by comparing the azimuthally-averaged reduced
shear profile about a chosen center of mass with reduced shear profiles calculated from various
mass models.  Provided the correct center of mass is chosen, tests with N-body simulations of
clusters have proved that this technique measures the correct spherically-averaged cluster mass 
profile on average, but can produce up to a $\sim 30\%$ error in the total mass of a given
cluster depending primarily on how closely aligned the major-axis of 3-D cluster mass
distribution is to the line of sight \citep{CL04.1}.   

To create the shear profiles, we azimuthally averaged the tangential shear measurements,
weighted by $\min (\nu_\mathrm{t}, 40)$, within logarithmically-spaced radial bins. {
The model shear profiles are also geometrically averaged within the same radial bins during
the fitting process.}  The error
in the shear measurements were calculated from the rms of the $45^\circ $ component
to the tangential shear of all of the galaxies and divided by the square root of the number
of galaxies in each bin.  The rms dispersion of the $45^\circ$ component for each cluster is
given in Tables \ref{lztable} and \ref{hztable}.

We tried fitting three different mass models to the clusters - a singular isothermal sphere
\begin{equation}
\kappa=\gamma_{\mathrm{t}}=2\pi {\sigma\over c}^2 {D_ls/D_os},
\end{equation}
a NFW profile \citep{NA97.6}, and a cored King profile \citep{CL04.2}.  We found that due to the
small field size of the optical images ($~7\arcmin$), the two parameters in both the NFW and
King profiles were extremely degenerate with each other, and thus could not be used to provide
an accurate mass profile for the cluster.  As such we only give the best-fit parameters and
errors on the SIS profile in Tables \ref{lztable} and \ref{hztable}.  We also quote an overall 
significance for the fit,
as determined by the difference in the $\chi^2$ of the best fit model and a 0 km/s model.
This significance provides a minimum value for the overall significance of the mass
detection, as it assumes that the SIS profile provides a correct description of the true
mass profile and that there are no other mass peaks in the fitting region.

In order to convert the mass models into reduced shear profiles, we first had to change the
surface density profiles into $\kappa$ profiles, which requires the knowledge of the mean
redshift of the background galaxies used to measure the reduced shear.  Because the background
galaxies used in the shear catalogs are too faint, we could not obtain either accurate 
photometric or spectroscopic redshifts.  Instead
we used the photometric redshift catalog of \citet{FO99.1} from the HDF-S, which has
the advantage of having photometry measured from the VLT, and therefore not needing any
photometric transformations of passbands (and associated systematic errors).  We quote in
Tables \ref{lztable} and \ref{hztable} the mean lensing redshifts for each of the fields, which 
were calculated by applying
the same magnitude and color cuts to the HDF-S catalog as were used for the background galaxy
catalog, and averaging the $\Sigma_{\mathrm{crit}}^{-1}(z_\mathrm{cl},z_\mathrm{bg})$ values
from each surviving galaxy.  Errors in the mean lensing redshifts were calculated from bootstrap
resamplings of the catalog.

The third method used is aperture densitometry \citep{FA94.1,CL00.1}, which,
like the shear profile fitting technique, produces a radial mass measurement from an arbitrarily
chosen center of mass:
\begin{eqnarray}
\zeta _\mathrm{c}(r_1,r_2,r_\mathrm{m}) & = & \bar{\kappa }(<r_1) - \bar{\kappa }(r_2<r<r_\mathrm{m}) \\
\nonumber & = & 2\int_{r_1}^{r_2}d\ln r \langle \gamma_\mathrm{T}\rangle + {2\over (1-r_2^2/r_\mathrm{m}^2)}
\int_{r_2}^{r_\mathrm{m}}d\ln r \langle \gamma_\mathrm{T}\rangle
\end{eqnarray}
where $r_2$ and $r_\mathrm{m}$ are the inner and outer radii of an annular region whose mean 
surface density is subtracted from that inside the cylinder of radius $r_1$, and
$\gamma_\mathrm{T}$ is the component of the shear for the galaxy at radius $r$ which is
tangential to the chosen center of mass.  Due to the subtraction of the mean $\kappa$ inside
the defined annular region, $\zeta_\mathrm{c}$ measures a lower bound for the mean surface
mass density inside $r_1$, and can be converted to a lower bound on the surface mass
by multiplying by $\pi r_1^2 \Sigma_{\mathrm{crit}}$.  Errors on $\zeta_\mathrm{c}$ were
calculated by assigning each galaxy an error on the shear equal to the rms of the
$45^\circ$ component of all of the galaxies in the field, and propagating the errors.
It is important to note that
the statistic assumes that one is measuring $\gamma$; because we must instead measure the
reduced shear $g$, the calculated surface masses will be slightly too large for regions
of high surface mass density $\kappa >0.1$.  This error, however, is much smaller than the 
random error for all of the fields in this paper.
The aperture densitometry profile for each cluster is shown in Figs.~2--21.  

\begin{table*}[t]
\caption{$z\sim 0.5$ clusters.  The columns show the cluster name, spectroscopic redshift, number density
of background galaxies, rms 1--D reduced shear measurement of a background galaxy, mean lensing
redshift of the background galaxies, best-fit SIS velocity dispersion with $1\sigma$ errors,
significance of the SIS fit, significance of the mass aperture peak, filter radius for the
mass aperture measurement, cluster mass-to-light ratio in $I$, and the cluster mass-to-light
ratio in $B$.}
\begin{tabular}{lcccccccccc}
cluster & $z_\mathrm{cl}$ & $n_\mathrm{g}$ ($\#/\Box \arcmin$) & $\sigma_g$ & 
$\bar{z}_{\mathrm{bg}}$ & $V$ (km/s) & $\sigma_V$ & $\sigma_\mathrm{map}$ &
$r_\mathrm{map}$ & $M/L_I$ & $M/L_B$\\ \hline
CLJ1018.8-1211 & 0.472 & 22.4 & 0.278 & 0.99 & $603^{+82}_{-100}$ & 3.2 & 2.9 & $2\farcm 0$ & $174^{+50}_{-53}$ & $235^{+68}_{-71}$\\
CLJ1059.2-1253 & 0.457 & 21.4 & 0.272 & 0.97 & $1033^{+69}_{-80}$ & 5.8 & 4.7 & $2\farcm 5$ & $339^{+47}_{-51}$ & $445^{+61}_{-66}$\\
CLJ1119.3-1129 & 0.550 & 30.9 & 0.283 & 1.03 & $349^{+142}_{-349}$ & 1.0 & 0.7 & $3\farcm 0$ & $137^{+134}_{-136}$ & $181^{+177}_{-181}$\\
CLJ1202.7-1224 & 0.424 & 25.6 & 0.308 & 0.96 & $447^{+136}_{-214}$ & 1.4 & 1.5 & $3\farcm 0$ & $191^{+134}_{-139}$ & $252^{+177}_{-184}$\\
CLJ1232.5-1250 & 0.542 & 35.4 & 0.289 & 1.03 & $948^{+50}_{-55}$ & 8.1 & 7.1 & $2\farcm 0$ & $189^{+20}_{-21}$ & $247^{+27}_{-28}$\\
CLJ1238.5-1144 & 0.465 & 34.2 & 0.295 & 0.98 & $375^{+138}_{-254}$ & 1.1 & 0.4 & $2\farcm 5$ & $116^{+101}_{-104}$ & $240^{+209}_{-215}$\\
CLJ1301.7-1139 & 0.485 & 36.4 & 0.277 & 0.99 & $628^{+86}_{-104}$ & 3.2 & 2.0 & $1\farcm 5$ & $142^{+42}_{-43}$ & $183^{+54}_{-56}$\\
CLJ1353.0-1137 & 0.577 & 33.8 & 0.272 & 1.05 & $546^{+130}_{-180}$ & 1.8 & 1.4 & $3\farcm 0$ & $107^{+57}_{-59}$ & $155^{+82}_{-85}$\\
CLJ1411.1-1148 &  0.52 & 34.4 & 0.302 & 1.01 & $594^{+103}_{-128}$ & 2.6 & 1.4 & $3\farcm 0$ & $132^{+50}_{-51}$ & $187^{+70}_{-62}$\\
CLJ1420.3-1236 & 0.497 & 24.7 & 0.241 & 1.00 & $582^{+98}_{-118}$ & 2.7 & 1.8 & $3\farcm 0$ & $125^{+45}_{-46}$ & $192^{+70}_{-70}$\\ \hline \hline
\end{tabular}
\label{lztable}
\end{table*}
\begin{table*}[t]
\caption{$z>0.6$ clusters.  Same columns as in Table \ref{lztable}}
\begin{tabular}{lcccccccccc}
cluster & $z_\mathrm{cl}$ & $n_\mathrm{g}$ ($\#/\Box \arcmin$) & $\sigma_g$ & 
$\bar{z}_{\mathrm{bg}}$ & $V$ (km/s) & $\sigma_V$ & $\sigma_\mathrm{map}$ &
$r_\mathrm{map}$ & $M/L_I$ & $M/L_B$\\ \hline
CLJ1037.9-1243 & 0.580 & 54.2 & 0.328 & 1.11 & $503^{+84}_{-102}$ & 2.7 & 3.0 & $2\farcm 5$ & $171^{+62}_{-62}$ & $183^{+66}_{-67}$\\
CLJ1040.7-1155 & 0.704 & 44.4 & 0.320 & 1.19 & $438^{+138}_{-221}$ & 1.3 & 1.0 & $2\farcm 5$ & $60^{+44}_{-46}$ & $74^{+54}_{-56}$\\
CLJ1054.4-1146 & 0.697 & 38.4 & 0.342 & 1.18 & $885^{+99}_{-117}$ & 3.8 & 2.9 & $1\farcm 0$ & $105^{+25}_{-26}$ & $151^{+36}_{-37}$\\
CLJ1054.7-1245 & 0.750 & 35.8 & 0.333 & 1.22 & $906^{+88}_{-102}$ & 4.4 & 4.9 & $1\farcm 5$ & $173^{+35}_{-37}$ & $217^{+44}_{-46}$\\
CLJ1103.7-1245 & 0.960 & 32.5 & 0.369 & 1.36 & $899^{+129}_{-159}$ & 3.0 & 2.9 & $2\farcm 5$ & $785^{+241}_{-253}$ & $1063^{+327}_{-343}$ \\
CLJ1122.9-1136 & 0.807 & 34.9 & 0.341 & 1.26 & $589^{+172}_{-262}$ & 1.4 & 2.3 & $1\farcm 0$ & $179^{+120}_{-124}$ & $208^{+139}_{-144}$\\
CLJ1138.2-1133 & 0.480 & 43.1 & 0.318 & 1.13 & $529^{+78}_{-96}$ & 2.9 & 1.5 & $1\farcm 0$ & $177^{+56}_{-58}$ & $209^{+66}_{-69}$\\
CLJ1216.8-1201 & 0.794 & 37.6 & 0.337 & 1.25 & $1152^{+70}_{-78}$ & 6.8 & 5.2 & $3\farcm 0$ & $98^{+12}_{-12}$ & $130^{+16}_{-17}$\\
CLJ1227.9-1138 & 0.634 & 31.1 & 0.326 & 1.14 & $308^{+221}_{-308}$ & 0.5 & 1.2 & $1\farcm 0$ & $51^{+100}_{-51}$ & $65^{+128}_{-65}$\\
CLJ1354.2-1230 & 0.757 & 37.1 & 0.299 & 1.23 & $747^{+87}_{-98}$ & 3.9 & 3.7 & $1\farcm 0$ & $137^{+34}_{-34}$ & $160^{+39}_{-39}$\\ \hline
\end{tabular}
\label{hztable}
\end{table*}

We also calculated the significance of each cluster detection using a 
mass aperture statistic \citep{SC96.3, SC98.3}:
\begin{equation}
M_{\mathrm{ap}} = 4 \sum_{i=1}^{n}\,x_i^2\,(1-x_i^2)\,g_{\mathrm{T,}i}
\end{equation}
where $x$ is the radial distance from galaxy $i$ to the aperture center
normalized by the maximum filter radius, $g_{\mathrm{T,}i}$ is the
component of the reduced shear measurement for galaxy $i$ tangential to
the center of the aperture, and the sum is taken over all galaxies within
the maximum radius of the aperture.  This statistic calculates the value of
the surface mass inside a chosen radius convolved with a compensated filter.
In the absence of a massive structure, the mass aperture statistic will have
a 0 average and a Gaussian error distribution from the intrinsic ellipticities
of the background galaxies.  The significance of the mass aperture statistic
therefore can be calculated by dividing the measured value by the rms value
of simulations performed by subtracting the azimuthally-averaged shear profile
about the chosen center from the background galaxy ellipticities, randomizing the
orientation of each background galaxy while preserving its position and total
shear, and calculating the mass aperture statistic on the randomized shears.
Because the filter is designed to be compensated in mass, however, it will
underestimate the significance of structures which are more extended than
the filter size as well as those with neighboring mass peaks.

We calculated the mass aperture statistic for a series of filter sizes
($0\farcm 5$ to $3\farcm 0$ in steps of $0\farcm 5$), and give in Tables 
\ref{lztable} and \ref{hztable}
the maximum significance and corresponding filter size.  Due to the compensated filter,
the significances for this technique are a lower bound on the significance
of the lensing detection, especially for clusters which are less
concentrated than average.  A small filter size for the maximum
significance generally indicates the presence of another large mass peak
in the field of view, although the higher redshift clusters will naturally have
a smaller filter size (as measured in arcminutes) due to the increasing angular 
diameter distance to the cluster.  

{For all three of the one-dimensional mass measurement techniques discussed above,
we assumed that the center of mass of the cluster was at the location of the
brightest cluster galaxy (hereafter BCG).  While there is often a small offset 
between the location of the BCG and the peak of the mass distribution in the
two-dimensional $\kappa$ maps, these offsets are usually consistent with being 
caused by noise peaks superimposed on the mass distribution \citep{CL00.1}.
Centering the one-dimensional mass measurements on these peaks in the $\kappa$ maps
results in a systematic overestimate of the mass and significance in all three
techniques described above \citep{VW00.1}.  If the observed offsets
are real, then using the BCG as the center results in a systematic underestimate
of the mass and significance measurements.  We therefore present the results in Tables
\ref{lztable} and \ref{hztable} with the caveat that they are minimum masses for
the clusters and are accurate only if the center of mass of the clusters is
spatially coincident with the BCG.}

\subsection{Mass-to-Light Ratios}

Likely cluster galaxies were selected from a union of two photometric redshift catalogs
using a technique described in \citet{PE04.1} which is based on the photometric redshift
code of \citet{BO00.1,RU01.1,RU03.1}.  In addition to the photometric redshift
selection, objects were only allowed into the cluster galaxy catalog if they had a
total $I$ flux and central surface brightness less than the brightest cluster
galaxy, whose selection is described in \citet{WH04.1}.  Absolute rest-frame luminosities
were then calculated from the cluster galaxy catalogs, taking into account Galactic
extinction \citep{SC98.1}, field crowding, and an aperture correction for the extended
PSF wings.  The details of the luminosity calculations can be found in \citet{RU04.1}.

The luminosity density of a cluster was determined by summing the absolute luminosities of
all of the cluster galaxies within 500 kpc of the BCG, dividing by the surface area
of the 500 kpc radius aperture, and subtracting the luminosity density of cluster galaxies
at distances greater than 1 Mpc from the BCG.  This last step corrects
for the fact that the photometric redshift selection can still allow a significant fraction
of interlopers among the true cluster galaxies in the catalog.  

The cluster mass density
was then measured from the best-fit SIS profile by taking the mean surface density within
500 kpc of the BCG and subtracting the mean surface density of those regions within the
image greater than 1 Mpc from the BCG.  The resulting mass-to-light ratios are given in
Tables \ref{lztable} and \ref{hztable} for both the rest frame $B$ and $I$ bands, and are
correct if the clusters have a constant mass-to-light ratio over the VLT images.  If the
mass-to-light ratio of the clusters is not a constant, then the measured values will have
a slight bias due to the subtraction of the large-radius luminosity and mass surface
densities.

\begin{figure*}
\sidecaption
\includegraphics[width=12cm]{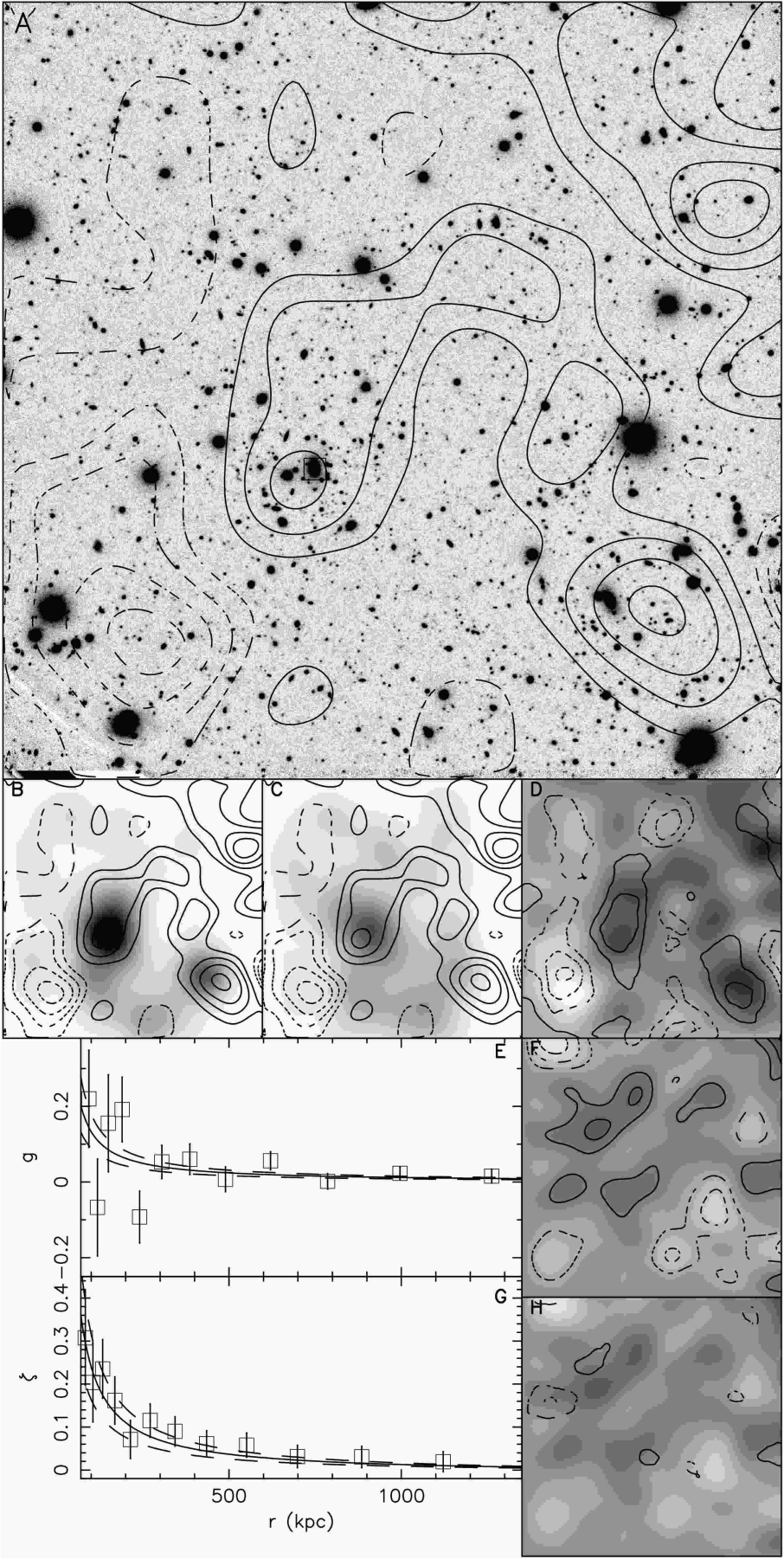}
\caption{A.  Shown in greyscale is the $I$-band image of CLJ1018.8$-$1211, $6\farcm 93$ or
2.45 Mpc in the cluster restframe on a side.  Overlayed in contours
are the weak lensing mass reconstruction of the field, smoothed with a $25\farcs 6$ Gaussian,
 with contour intervals of $10^8 \mathrm{M}_\odot/\mathrm{kpc}^2$.  The solid lines indicate 
positive mass and the dashed contours indicate negative mass relative to the mean mass 
density at the edge of the image.  A box is drawn around the chosen BCG, which is used as the
center of the cluster for the radial profile fits.  In this an all other figures north is
up and east is to the left.
B.  Shown in greyscale is the luminosity density of cluster 
galaxies,
as selected by photometric redshifts, smoothed by the same amount as the mass reconstruction.
Overlayed in contours is the mass reconstruction, with the same contour levels as in panel A.
C.  Shown in greyscale is the number density of cluster galaxies, with the same smoothing
and contours as panel B.
D. Shown in greyscale is the mass reconstruction of the field.  Overlayed in contours are 
significance levels of the peaks as calculated from
the mass aperture statistic, with each contour giving a change of $1\sigma$ significance,
and calculated from a $2\farcm 5$ aperture radius.
E.  The data points (with error bars) show the azimuthally-averaged
reduced shear profile for the cluster, centered on the brightest cluster galaxy.  The solid
line gives the reduced shear profile for the best fitting SIS model, and the dashed lines show
the reduced shear profiles for the $1\sigma$ variations in the fit models.
F.  Shown in greyscale is a mass reconstruction of the field after all of the background galaxies
were rotated by $45^\circ$, using the same greyscale as in panel D.  This provides a good assessment 
of the noise level in the reconstruction.  Overlayed are contours of this reconstruction, with the 
same mass density levels as shown in panel A.  
G.  The data points show the aperture densitometry profile of the cluster calculated from
the shear profile in panel E.  The solid line gives the SIS profile of the model which was
best-fit to the shear profile, with the dashed lines indicating the $1\sigma$ variations.
H. Shown in greyscale is the same noise reconstruction as panel F.  Overlayed are contours of the
mass aperture statistic calculated from the $45 ^\circ$ rotated background galaxies, using
the same contour levels and mass aperture radius as in panel D.
}
\label{cl1018}
\end{figure*}

\begin{figure*}
\sidecaption
\includegraphics[width=12cm]{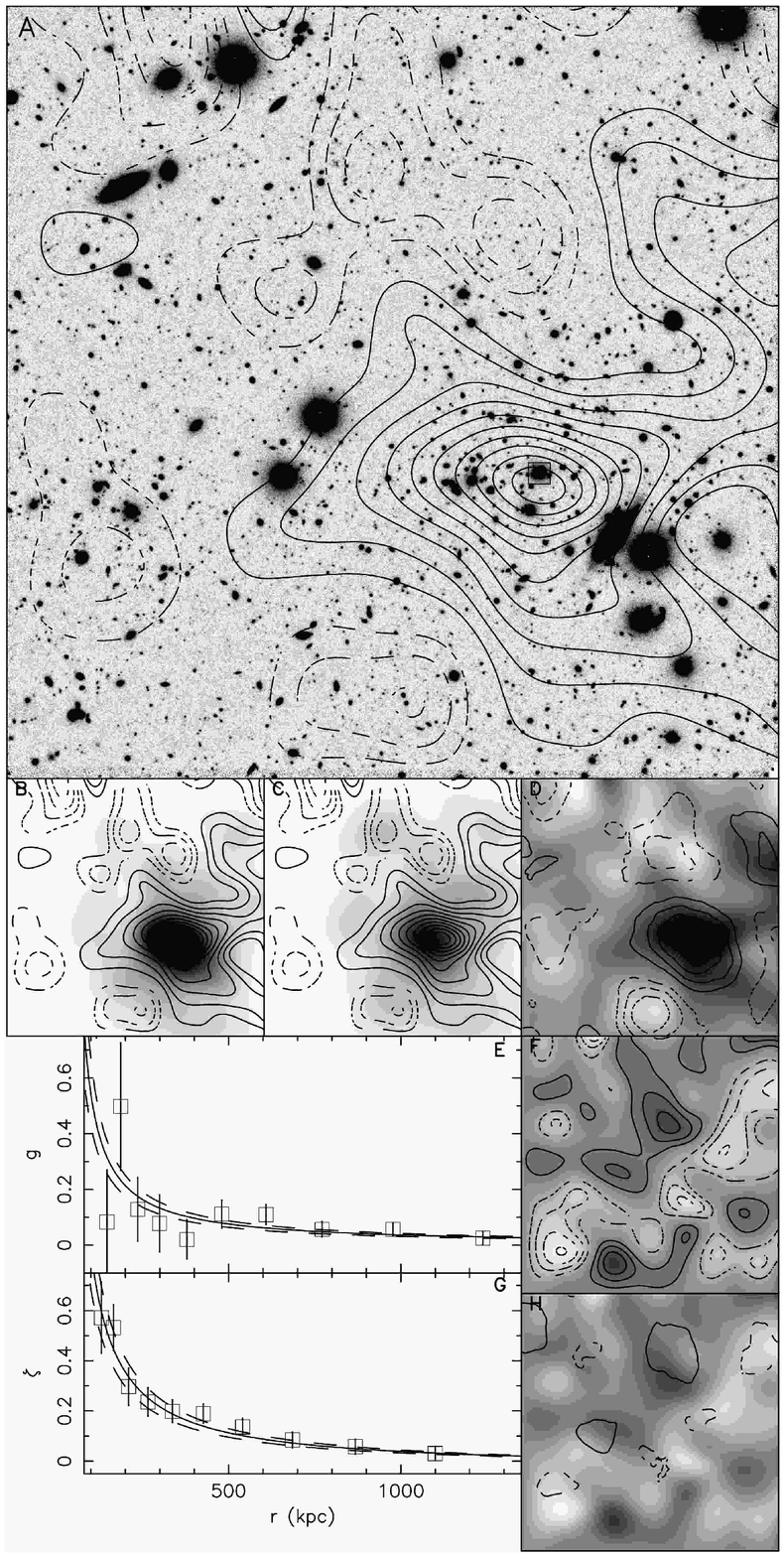}
\caption{Shown in panels A--F are the figure for CLJ1059.2$-$1253, with images
  $6\farcm 93$ or 2.40 Mpc in the cluster restframe on a side, using the same
  layout as Fig.~\ref{cl1018}.  The aperture densitometry significance
  contours of panels D and H use a $2\farcm 5$ aperture radius.}
\label{cl1059}
\end{figure*}

\begin{figure*}
\sidecaption
\includegraphics[width=12cm]{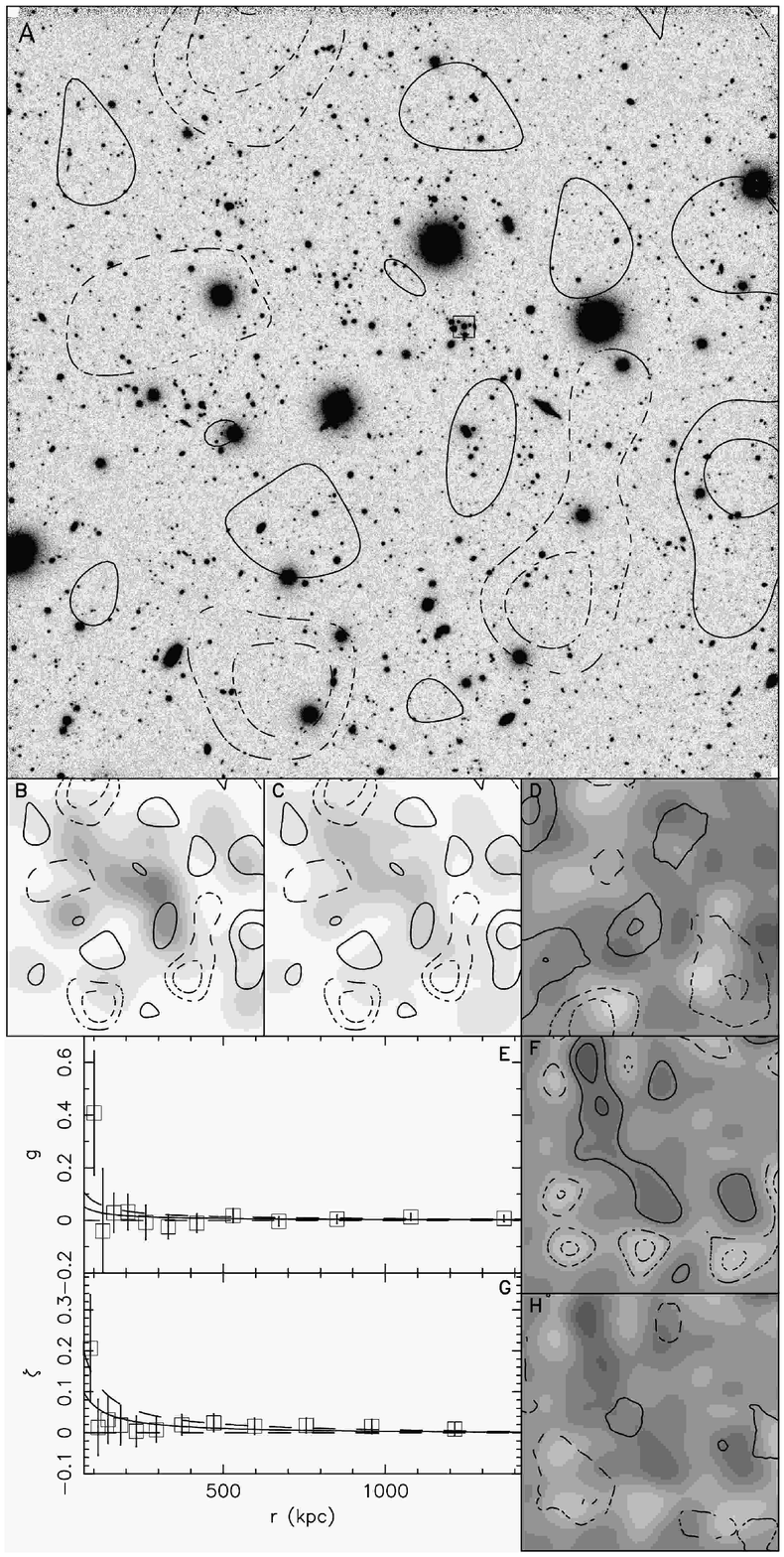}
\caption{Shown in panels A--F are the figure for CLJ1119.3$-$1129, with images
  $6\farcm 99$ or 2.69 Mpc in the cluster restframe on a side, using the same
  layout as Fig.~\ref{cl1018}.  The aperture densitometry significance
  contours of panels D and H use a $3\farcm 0$ aperture radius.}
\label{cl1119}
\end{figure*}

\begin{figure*}
\sidecaption
\includegraphics[width=12cm]{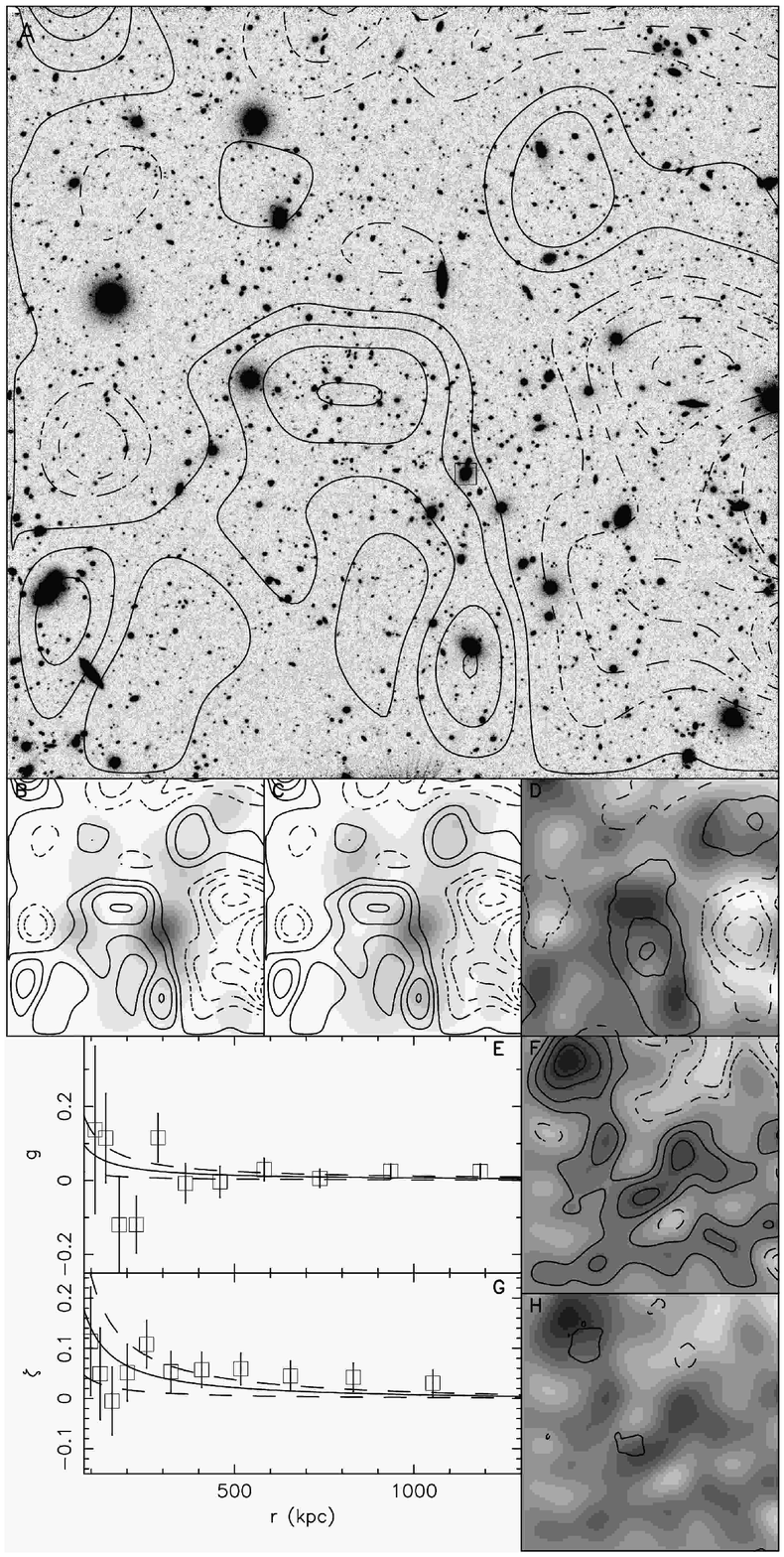}
\caption{Shown in panels A--F are the figure for CLJ1202.7$-$1224, with images
  $6\farcm 99$ or 2.32 Mpc in the cluster restframe on a side, using the same
  layout as Fig.~\ref{cl1018}.  The aperture densitometry significance
  contours of panels D and H use a $3\farcm 0$ aperture radius.}
\label{cl1202}
\end{figure*}

\begin{figure*}
\sidecaption
\includegraphics[width=12cm]{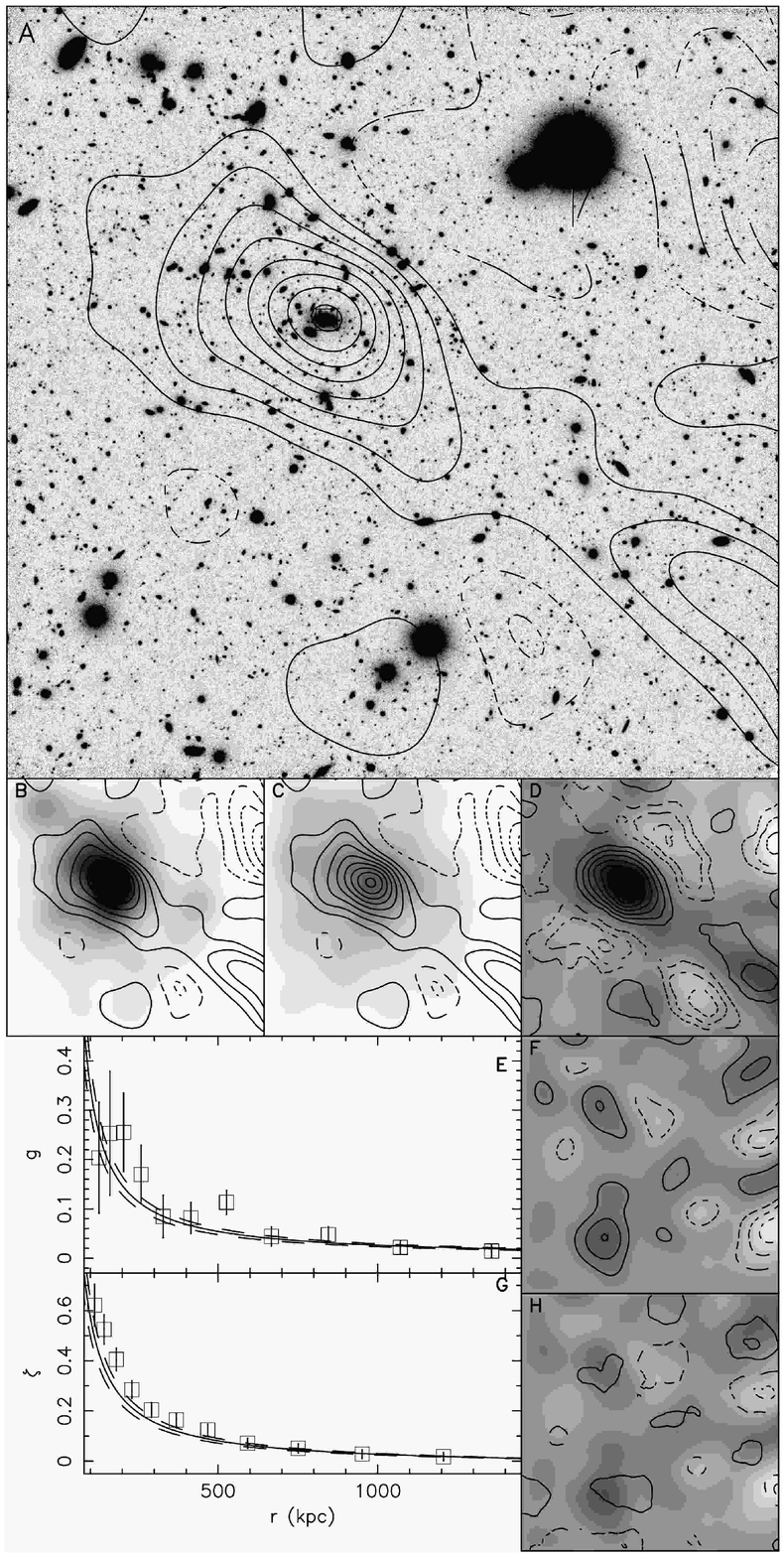}
\caption{Shown in panels A--F are the figure for CLJ1232.5$-$1250, with images
  $6\farcm 99$ or 2.66 Mpc in the cluster restframe on a side, using the same
  layout as Fig.~\ref{cl1018}.  The greyscale in panels B and C have their
  maximum value at twice that in Fig.~\ref{cl1018}.  The aperture densitometry significance
  contours of panels D and H use a $2\farcm 0$ aperture radius.}
\label{cl1232}
\end{figure*}

\begin{figure*}
\sidecaption
\includegraphics[width=12cm]{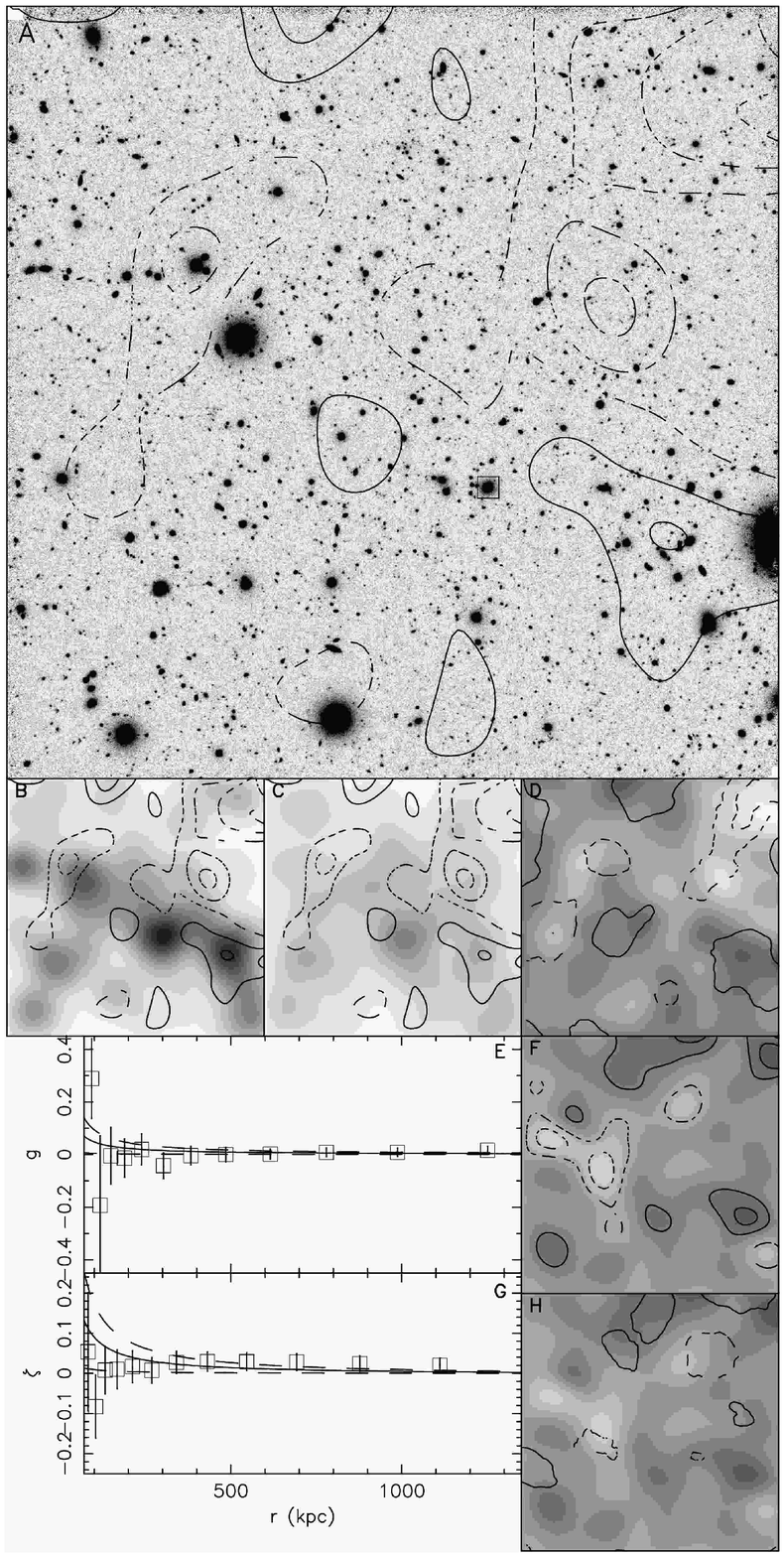}
\caption{Shown in panels A--F are the figure for CLJ1238.5$-$1144, with images
  $6\farcm 99$ or 2.44 Mpc in the cluster restframe on a side, using the same
  layout as Fig.~\ref{cl1018}.  The aperture densitometry significance
  contours of panels D and H use a $2\farcm 5$ aperture radius.}
\label{cl1238}
\end{figure*}

\begin{figure*}
\sidecaption
\includegraphics[width=12cm]{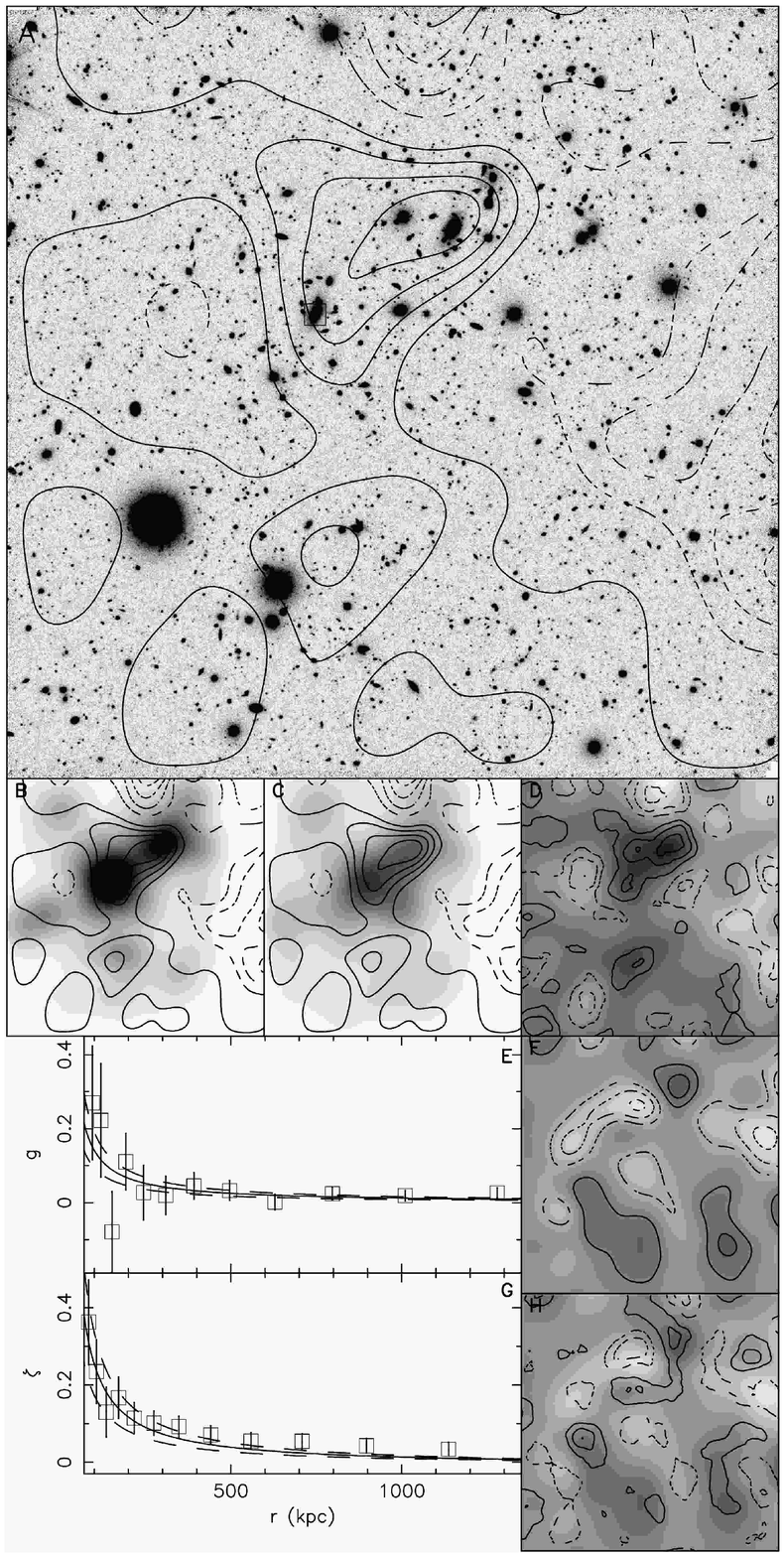}
\caption{Shown in panels A--F are the figure for CLJ1301.7$-$1139, with images
  $6\farcm 99$ or 2.50 Mpc in the cluster restframe on a side, using the same
  layout as Fig.~\ref{cl1018}.  The aperture densitometry significance
  contours of panels D and H use a $3\farcm 0$ aperture radius.}
\label{cl1301}
\end{figure*}

\begin{figure*}
\sidecaption
\includegraphics[width=12cm]{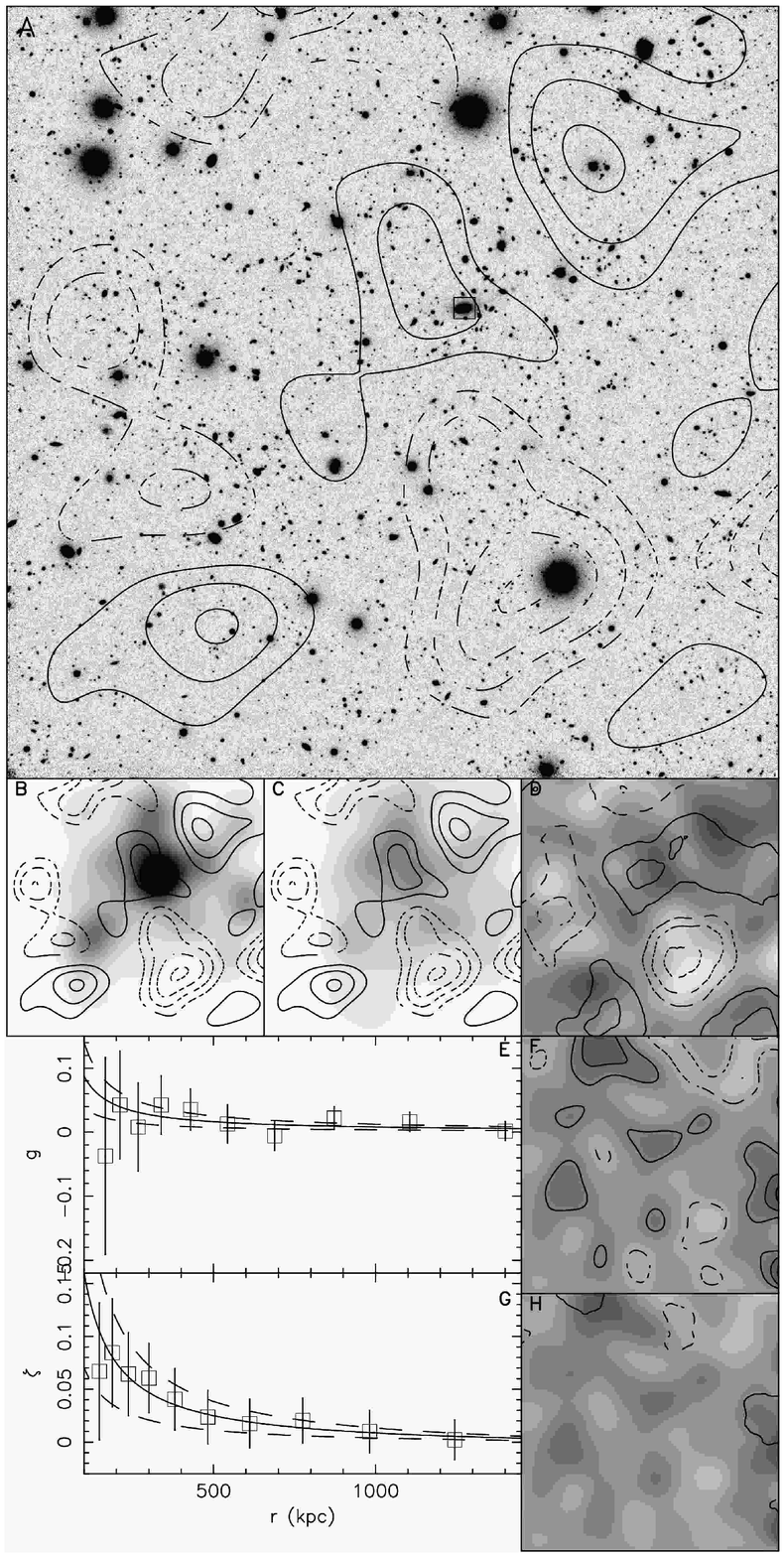}
\caption{Shown in panels A--F are the figure for CLJ1353$-$1137, with images
  $6\farcm 93$ or 2.76 Mpc in the cluster restframe on a side, using the same
  layout as Fig.~\ref{cl1018}.  The aperture densitometry significance
  contours of panels D and H use a $3\farcm 0$ aperture radius.}
\label{cl1353}
\end{figure*}

\begin{figure*}
\sidecaption
\includegraphics[width=12cm]{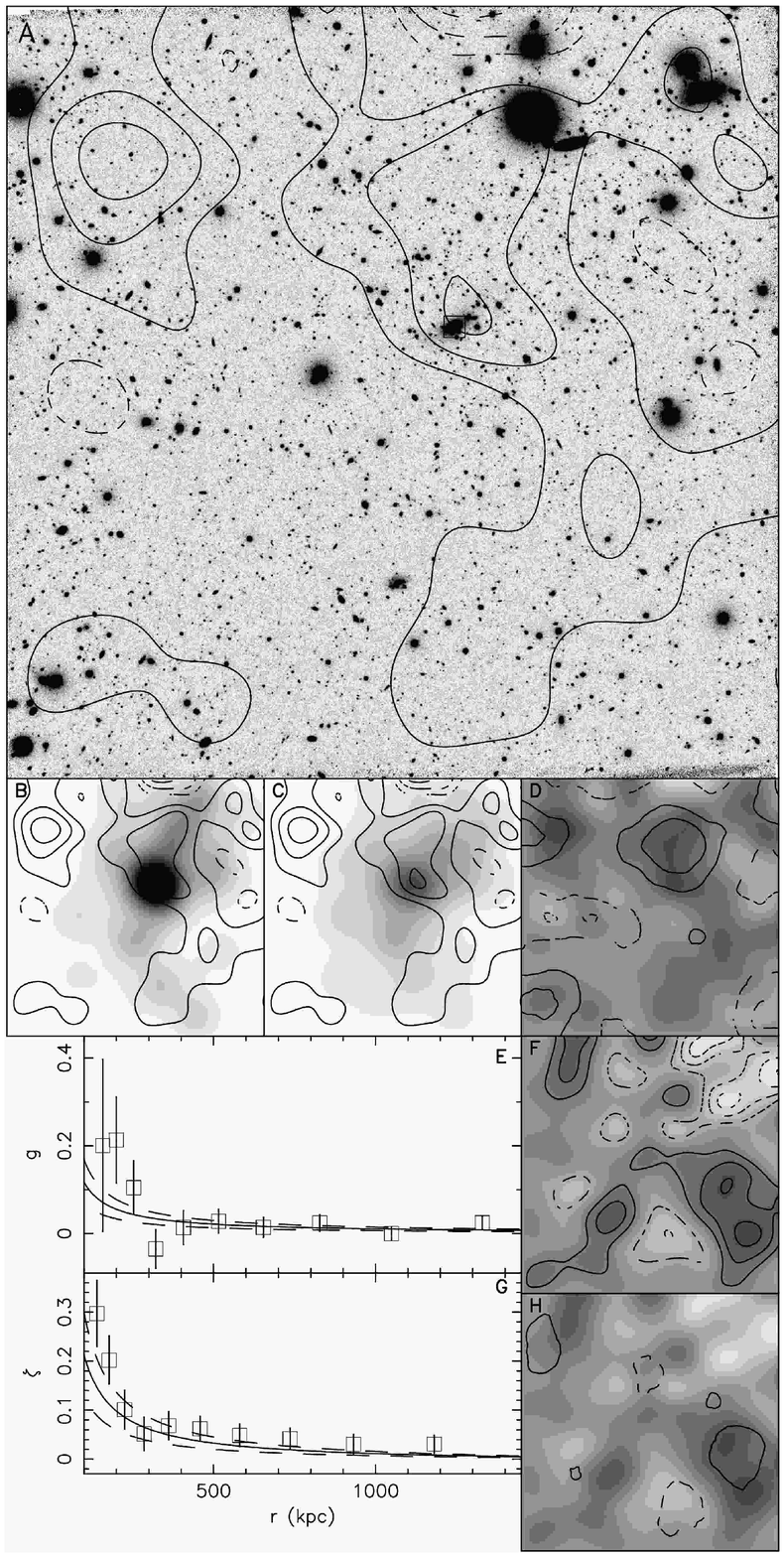}
\caption{Shown in panels A--F are the figure for CLJ1411.1$-$1148, with images
  $6\farcm 99$ or 2.61 Mpc in the cluster restframe on a side, using the same
  layout as Fig.~\ref{cl1018}.  The aperture densitometry significance
  contours of panels D and H use a $3\farcm 0$ aperture radius.}
\label{cl1411}
\end{figure*}

\begin{figure*}
\sidecaption
\includegraphics[width=12cm]{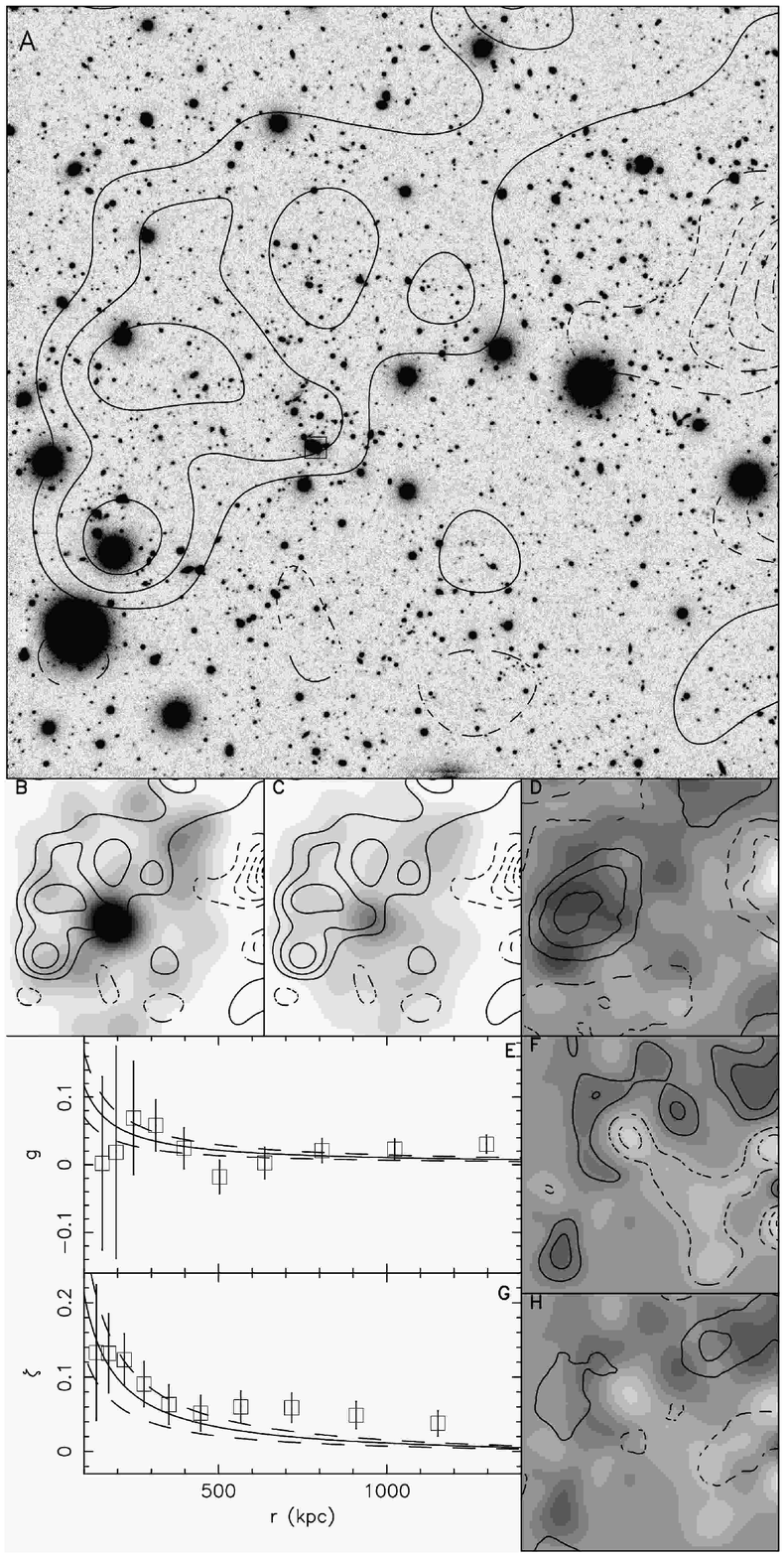}
\caption{Shown in panels A--F are the figure for CLJ1420.3$-$1236, with images
  $6\farcm 6$ or 2.44 Mpc in the cluster restframe on a side, using the same
  layout as Fig.~\ref{cl1018}.  The aperture densitometry significance
  contours of panels D and H use a $3\farcm 0$ aperture radius.}
\label{cl1420}
\end{figure*}

\section{$z\sim 0.5$ Clusters}

The weak lensing analyzes for the intermediate redshift sample are shown in Figs.~2--11, all of which
have the same layout.  The top panel (A) shows the $I$-band VLT image in greyscale overlayed with
contours of the mass reconstruction.  Each contour represents a change in the surface mass
density of $10^8 \mathrm{M}_\odot/\mathrm{kpc}^2$ relative to the mean surface mass at the
edge of the image, with solid contours indicating a surface mass increase and dashed contours
indicating a decrease.  The mass reconstruction has been smoothed with a $25\farcs 6$ circular
Gaussian profile to remove high-frequency noise, which results in the cluster mass peaks having
a broad core in the reconstruction even if the true mass profile has a cusp.
As the edges of an image are located typically between 1000 kpc (nearest
edge) and 1800 kpc (furthest corner) from the cluster core, the mean surface mass at the edge
will still be significantly above the cosmic mean, and thus some of the negative mass contours
may be real.  The mass reconstructions are, to a good approximation, the sum of the true
surface mass density field and a white-noise field resulting from the intrinsic ellipticities
of the background galaxies.  This noise field results not only in false mass peaks, but also
in the displacement of true mass peaks, with less massive peaks suffering, on average, a greater
displacement.  Simulations have shown that a $5\sigma$ peak can be routinely displaced by up
to $10\arcsec$ from its true position \citep{CL00.1}.  The chosen BCG for each cluster is
indicated by a box drawn around it.  In all of the figures north is up and east is to the left.

The middle left-hand panel (B) shows a luminosity map, also smoothed with a $25\farcs 6$ circular
Gaussian profile, of the photometric-redshift selected
cluster galaxies in greyscale with the positive mass weak lensing reconstruction contours 
overlayed.  The middle panel (C) shows a greyscale map of the number density of the
cluster galaxies, again smoothed with a $25\farcs 6$ circular Gaussian profile, with the
positive mass weak lensing reconstruction contours overlayed.  The greyscale for all of the clusters
in panels B and C are identical except for CLJ1232.5$-$1250, which has the maximum of the scale
extending to twice the values of the other clusters.  
The upper-middle right-hand panel (D) shows the mass reconstruction in greyscale with significance
contours of the mass aperture statistic overlayed, with each contour indicating a significance
step of $1\sigma$.  {The difference in positions of the reconstruction mass peaks and the
mass aperture significance peaks are a result of the different weighting functions the two
methods give to the shear field, and can be used as an estimate to the amount of error in the
centroid positioning of the mass peaks due to noise.  Using the mass aperture significance
map to measure the significance of additional peaks results in a systematic overestimate
of the peaks' significance \citep{VW00.1} as what is being measured is the significance of
the highest noise peak superimposed on the underlying mass distribution.}

The two graphs show the reduced shear profile about the
BCG (panel E) and the resulting aperture densitometry profile (panel G) as a function of
radius from the BCG.  Both graphs show the best fitting SIS profile (solid line) to the
reduced shear profile and the $1\sigma$ errors (dashed lines).

The lower-middle right-hand panel (F)
shows in both greyscale and contours, using the same contour levels as the mass reconstruction
in the previous panels, the mass reconstruction of the field after rotating all of the background
galaxies by 45 degrees.  { The mean shear profile of the cluster was first subtracted from
the background galaxies' shear measurements prior to the rotation.}  
Because the weak lensing shear field is irrotational, the rotation of
the background galaxies (equivalent to taking the curl of the shear field) produces an estimate
of the noise in the reconstruction.  The lower right-hand panel (H) has the same noise field in
greyscale as panel F, but has the significance of the mass aperture statistic for the
45 degree rotated background galaxies overlayed in contours, with each contour indicating
a significance step of $1\sigma$.

A summary of the weak lensing data for each cluster can be found in Table \ref{lztable}.
A description of each cluster's lensing analysis follows.

\subsection{CLJ1018.8$-$1211}

CLJ1018.8$-$1211 ($z = 0.472$) is detected as a $\sigma_\mathrm{v} = 603^{+82}_{-100}$ km/s
cluster at moderate significance ($3.2\sigma$ in shear, $2.9\sigma$ in $M_{\mathrm{ap}}$).
The offset observed between the BCG and the mass peak in the reconstruction is consistent
with shifts of peaks due to noise with this significance.  A second peak, with higher $\kappa$ but lower significance,
is also detected in the south-west corner of the field.  No spectra currently exist for this
peak, but photometric redshifts indicate that is likely at a lower redshift and therefore
unrelated to this cluster.  The relatively high mass-to-light ratio of this cluster may
indicate that the lensing mass includes some significant contribution from the lower
redshift structure. { The structure in the northwest corner in the mass reconstructions
is detected at only $1\sigma$ in $M_{\mathrm{ap}}$, so is likely a noise peak.}

\subsection{CLJ1059.2$-$1253}

CLJ1059.2$-$1253 ($z = 0.457$) is the second most massive cluster in the sample, and the most
massive in the intermediate redshift subsample, having a best fit 
SIS profile with $\sigma_\mathrm{v} = 1033^{+69}_{-80}$ km/s and
is detected at $5.8\sigma$ with the shear profile fitting and $4.7\sigma$ with
$M_\mathrm{ap}$.  The relatively low significance for the detection of a cluster of this
mass is due to all three VLT passbands having higher than average seeing levels, and 
therefore having the lowest number density of background galaxies with which to measure
the shear field.  { The $2\sigma$ peak located northwest of the cluster at the edge of the
VLT field does not correspond to an overdensity of galaxies within the VLT image, but could,
in theory, be caused by a structure immediately outside of the image.}

\subsection{CLJ1119.3$-$1129}

No significant peaks are detected in this field.  The shear profile about the BCG does
indicate a positive mass, although only at $1.0\sigma$ significance.  A $2\sigma$ upper
limit on the velocity dispersion of the cluster is 600 km/s.

\subsection{CLJ1202.7$-$1224}

Several small peaks in this field exist in the mass reconstruction, the nearest of which
is $\sim 1\farcm 5$ away from the likely BCG.  In the $M_{\mathrm{ap}}$ map, however,
a significant peak is located near this BCG, about $\sim 30\arcsec$ to the southeast.
Using this galaxy ($z = 0.424$)
as the BCG gives a best fit SIS profile of $447^{+136}_{-214}$ km/s with a significance 
of $1.4\sigma$ from the shear profile and $1.5\sigma$ with $M_{\mathrm{ap}}$.  { Using a
slightly fainter galaxy, located $28\arcsec$ south-east of the BCG, as the center
of mass results in a best fit SIS profile of $570^{+120}_{-164}$ km/s with $2.0\sigma$
significance, $2.6\sigma$ significance with $M_{\mathrm{ap}}$.  We do not have redshift
information on this second galaxy, but photometric redshifts give it a high probability
of being at the same redshift as the BCG.}  Several of the other peaks, none of which
are significant at more than $1.3\sigma$, are also spatially coincident with red
galaxy overdensities.  The redshifts of these galaxy overdensities are not currently
known, so it is uncertain if these smaller mass peaks are physically associated
with the cluster.

\subsection{CLJ1232.5$-$1250}

CLJ1232.5-1250 ($z = 0.542$) is the third most massive cluster in the sample,
having a best fit SIS profile with $\sigma_\mathrm{v} = 948^{+50}_{-51}$ km/s.  The
cluster is detected at $8.1\sigma$ significance with the shear profile and $7.1\sigma$
with $M_{\mathrm{ap}}$, which makes it the most significant lensing detection in the EDisCS
sample.  The mass peak, as one would expect for a peak with this high significance,
is spatially coincident with the brightest cluster galaxy in both the mass reconstruction
and $M_{\mathrm{ap}}$.  There is a second mass peak in the south-west corner of the
image which is also detected at more than $3\sigma$ significance and is
near a group of red galaxies which have colors consistent with the cluster red-sequence.
This secondary peak is the reason the filter radius for the highest significance
$M_{\mathrm{ap}}$ statistic is smaller than most of the other clusters in the intermediate redshift
sample.  The larger radius would put this peak in the negative weight portion
of the filter, and therefore decrease the significance of the cluster detection.  As
such, the $7.1\sigma$ significance is only a lower bound on the true significance
of the cluster peak.  The weak lensing mass measurement is in reasonable agreement with
the $\sigma_\mathrm{v} = 1080^{+119}_{-89}$ km/s velocity dispersion measured
spectroscopically with 58 cluster galaxies \citep{HA04.1}.

\subsection{CLJ1238.5$-$1144}

No significant peaks are detected in the mass reconstruction of this field.  Using the
likely BCG as the center of the shear profile gives a best-fit velocity dispersion of
$\sigma_\mathrm{v} = 375^{+138}_{-254}$ km/s and a significance of $1.1\sigma$ from
the shear profile, but only a $0.4\sigma$ significance from $M_\mathrm{ap}$.

\subsection{CLJ1301.7$-$1139}

One massive peak, spatially coincident with several galaxy overdensities, is evident in
the mass reconstruction shown in Fig.~\ref{cl1301}.  At smaller smoothing radii, however,
this peak breaks up into 3 peaks, each one located near one of the bright galaxies
surrounding the peak shown in the figure.  The southeast galaxy overdensity is the
structure which was targeted by the observations and is a galaxy cluster at $z=0.48$.
The northwest structure appears to be a second galaxy cluster at $z=0.39$, and it is
currently unknown whether the middle structure is associated with one of these two
clusters.  These two clusters are too close in redshift to allow the photometric
redshift measurements with our filter set to accurately determine to which cluster 
a given galaxy belongs, so the luminosity measurement can only be
made for the sum of the two clusters.  Using the BCG of the $z=0.48$ cluster as the
center of the shear profile gives a best-fit velocity dispersion of $\sigma_\mathrm{v} =
628^{+86}_{-104}$ km/s.  This mass estimate, however, is certain to be contaminated with
mass from the lower redshift cluster.  The mass-to-light ratio for the system
is consistent with the sample mean, as both the mass and luminosity measurements
are being contaminated by the lower redshift cluster. { Attempts to simultaneously fit
both clusters result in a high degree of degeneracy between the two mass measurements.}

\subsection{CLJ1353.0$-$1137}

A $\sigma_\mathrm{v} = 546^{+130}_{-180}$ km/s peak, detected at $1.8\sigma$ significance
in the shear profile and $1.4\sigma$ significance with $M_{\mathrm{ap}}$, is spatially
coincident with the CLJ1353.0-1137 BCG ($z = 0.577$).  Another peak, with higher $\kappa$
but a lower significance, is located north-west of the cluster and between two
galaxy overdensities.  The southern-most of these galaxies overdensities has photometric
redshifts consistent with the cluster, while the northern overdensity appears to be at
lower redshift.  { The peak south-east of the cluster in the mass reconstruction is not
spatially coincident with any galaxy overdensities.  Neither of the additional peaks are
significant in $M_\mathrm{ap}$ measurements.}

\subsection{CLJ1411.1$-$1148}

A broad mass peak of low significance is detected at the location of the BCG ($z = 0.520$).
Using the BCG as the center of the peak gives a best fit SIS profile of $\sigma_\mathrm{v}
= 594^{+103}_{-128}$ km/s with a significance of $2.6\sigma$ from the shear profile and
$1.4\sigma$ with $M_{\mathrm{ap}}$.  The mass reconstruction shows evidence for long
extensions to the south and west, although again at low significance.  The extensions
are not detected using $M_{\mathrm{ap}}$, although this is not unexpected as the
cluster would be in the negative weight portion of the filter and therefore
canceling any possible filamentary mass in the positive weight portion.  Any filamentary
mass will also be in the negative weight portion of the filter when centered on the cluster,
which could explain the lower significance from $M_{\mathrm{ap}}$.  An additional, low
significance peaks are found in the north-east and south-east corners of the field, possibly associated with
a lower-redshift galaxy overdensities found to near the peaks.

\subsection{CLJ1420.3$-$1236}

A broad mass peak is detected at moderate significance ($3.7\sigma$ from shear, $3.2\sigma$
from $M_{\mathrm{ap}}$) near CLJ1420.3$-$1236 ($z = 0.497$). The offset between the centroid 
of the peak and the BCG is $\sim 50\arcsec$, which for a $4\sigma$ peak is significant at
$\sim 3\sigma$.  At smaller smoothing filters than that shown in Fig.~\ref{cl1420}, however,
the peak breaks into two components, one of which is $\sim 20\arcsec$ from the BCG.  The
second component is not located near any significant concentration of galaxies.  Using the
BCG as the center of the cluster, one obtains a shear profile which is best fit by
a $\sigma_\mathrm{v} = 583^{+98}_{-118}$ km/s velocity dispersion and is significant at 
$2.7\sigma$ in the shear profile $1.8\sigma$ in $M_{\mathrm{ap}}$.  Two additional peaks are
located $\sim 1\farcm 5$ north-west and south-east of the cluster, both near galaxy
overdensities which have the same photometric redshift as the cluster.  The multiple peaks
and the fact the reduced shear and aperture densitometry mass at large radius from the 
cluster core are much larger than the best fit SIS model predicts indicate that this 
is a massive cluster which is either just forming or undergoing one or more major mergers.

\section{$z\sim 0.7$ Clusters}

\begin{figure*}
\sidecaption
\includegraphics[width=12cm]{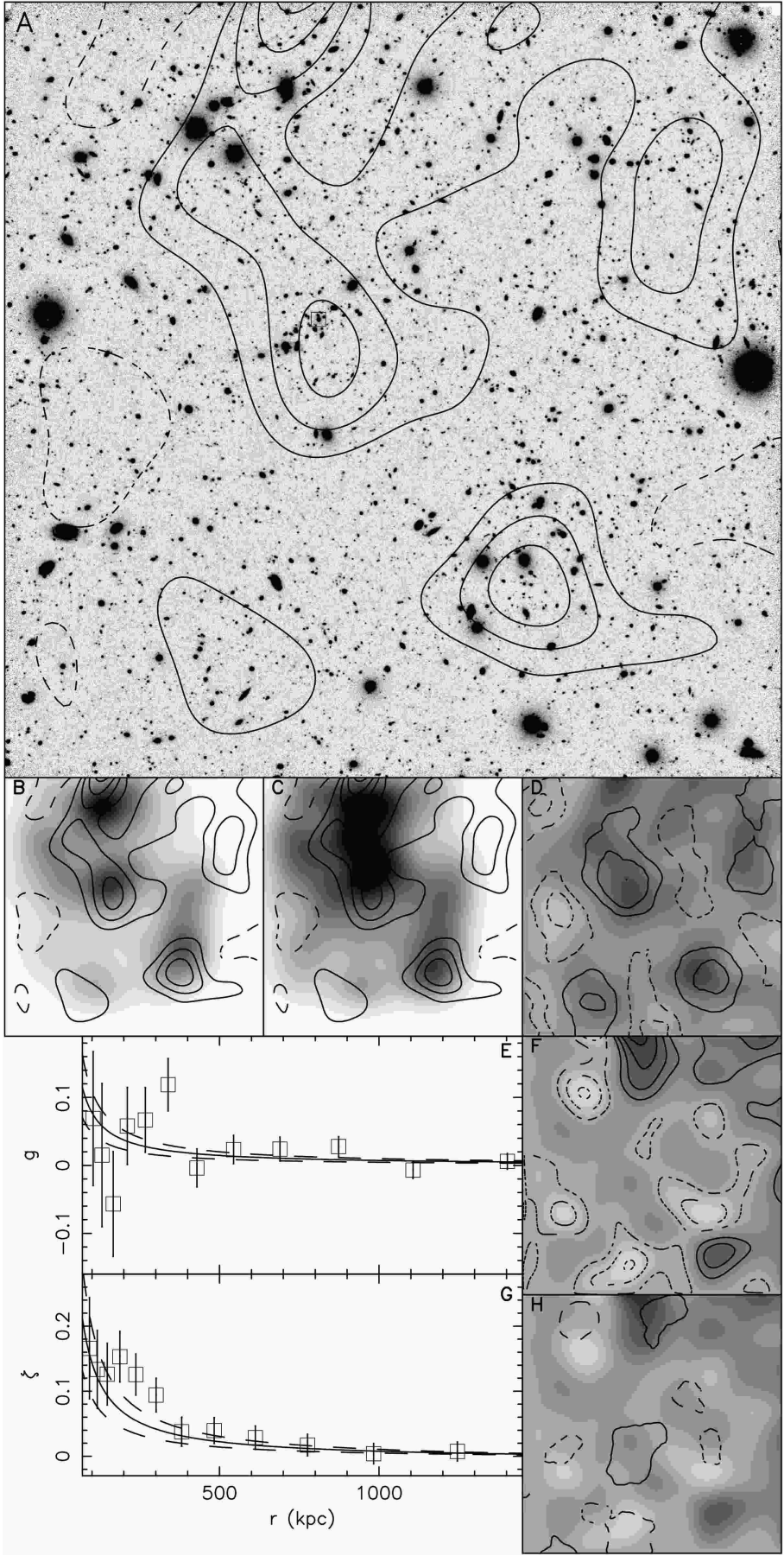}
\caption{Shown in panels A--F are the figure for CLJ1037.9$-$1243, with images
  $6\farcm 80$ or 2.80 Mpc in the cluster restframe on a side, using the same
  layout as Fig.~\ref{cl1018}.  The aperture densitometry significance
  contours of panels D and H use a $2\farcm 5$ aperture radius.}
\label{cl1037}
\end{figure*}

\begin{figure*}
\sidecaption
\includegraphics[width=12cm]{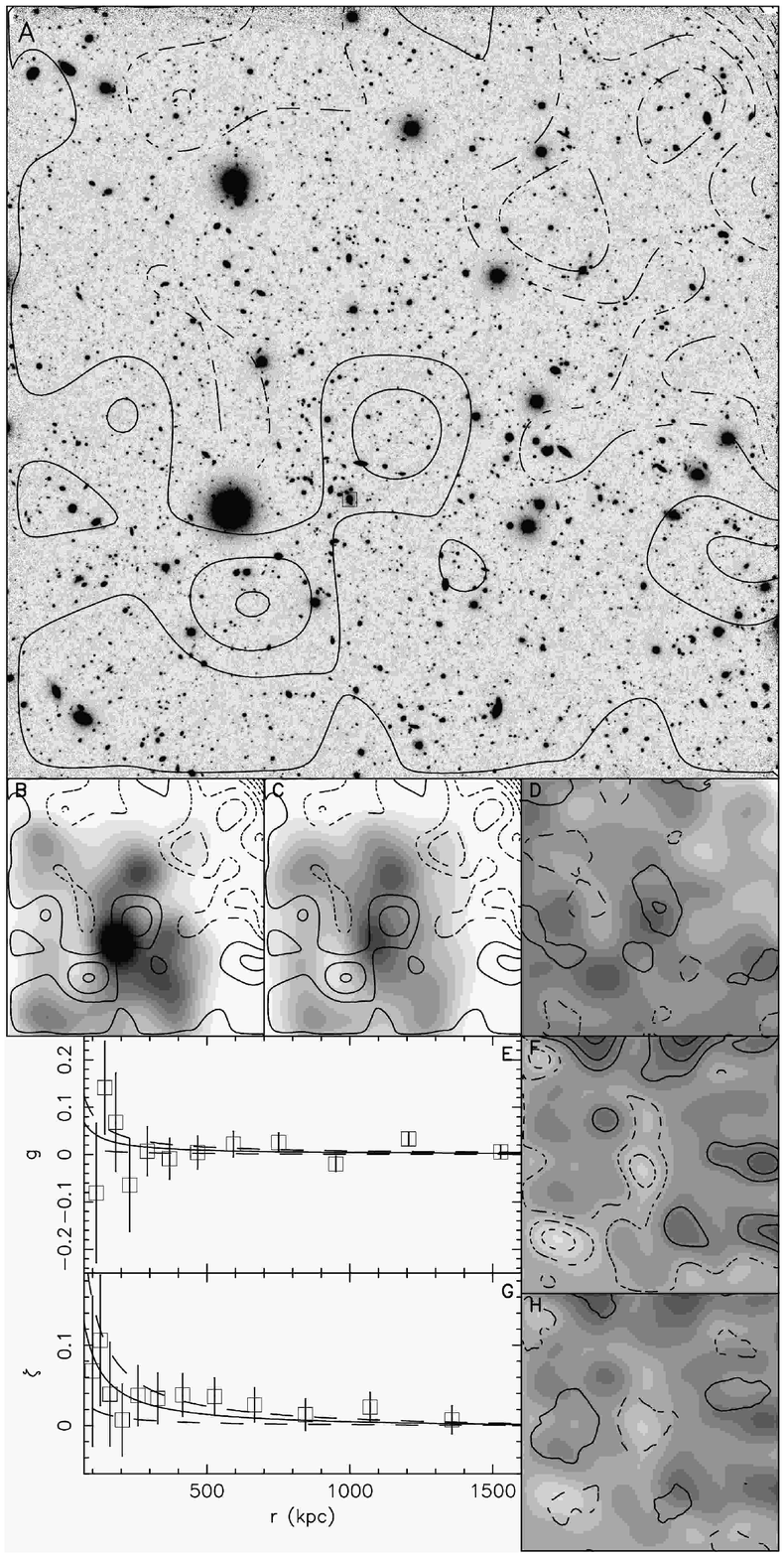}
\caption{Shown in panels A--F are the figure for CLJ1040.7$-$1155, with images
  $7\farcm 15$ or 3.07 Mpc in the cluster restframe on a side, using the same
  layout as Fig.~\ref{cl1018}.  The aperture densitometry significance
  contours of panels D and H use a $3\farcm 0$ aperture radius.}
\label{cl1040}
\end{figure*}

\begin{figure*}
\sidecaption
\includegraphics[width=12cm]{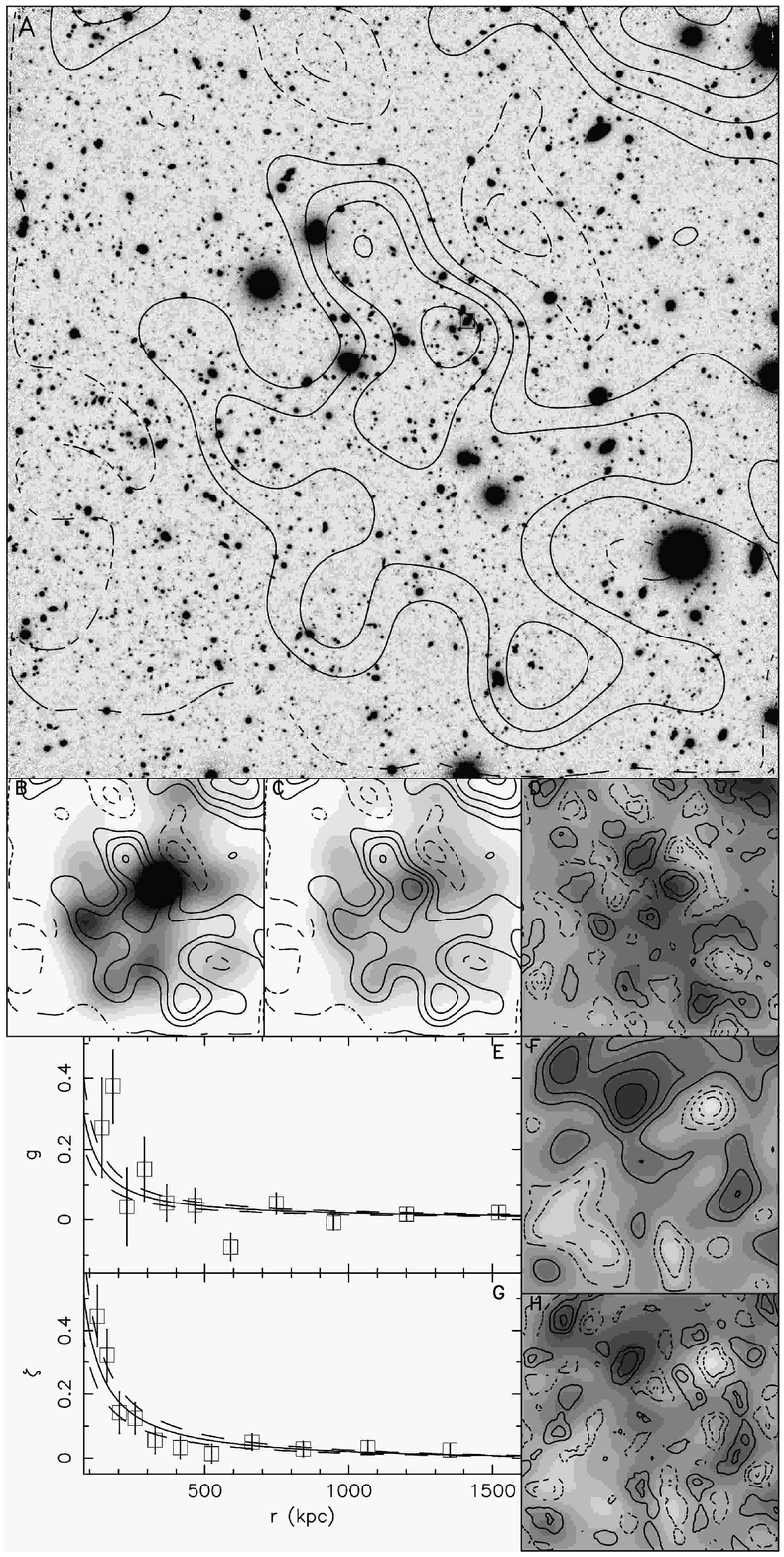}
\caption{Shown in panels A--F are the figure for CLJ1054.4$-$1146, with images
  $7\farcm 09$ or 3.04 Mpc in the cluster restframe on a side, using the same
  layout as Fig.~\ref{cl1018}.  The greyscale in panels B and C have their
  maximum value at twice that in Fig.~\ref{cl1018}.  The aperture densitometry significance
  contours of panels D and H use a $1\farcm 0$ aperture radius.}
\label{cl1054-11}
\end{figure*}

\begin{figure*}
\sidecaption
\includegraphics[width=12cm]{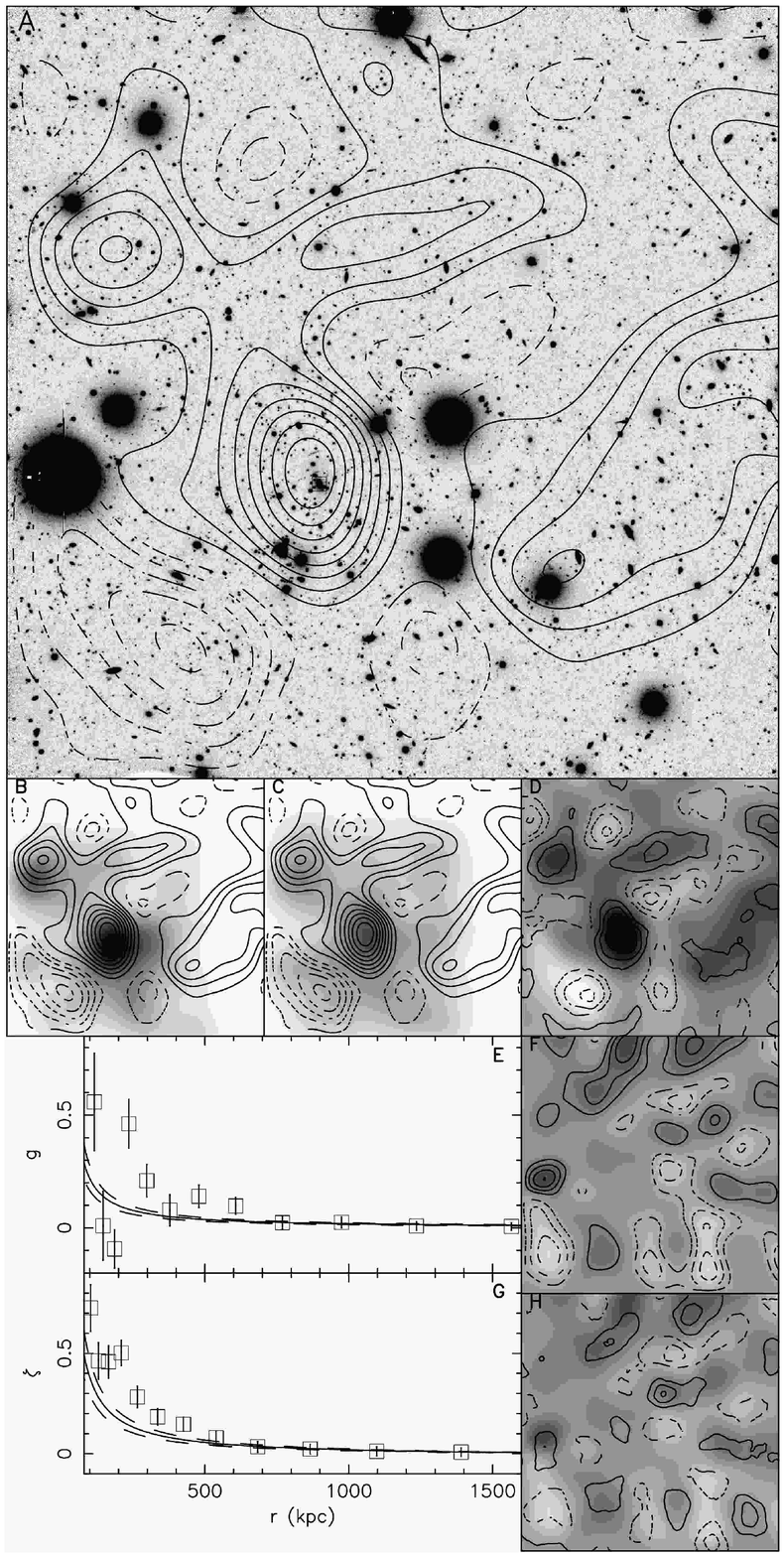}
\caption{Shown in panels A--F are the figure for CLJ1054.7$-$1245, with images
  $7\farcm 04$ or 3.10 Mpc in the cluster restframe on a side, using the same
  layout as Fig.~\ref{cl1018}.  The greyscale in panels B and C have their
  maximum value at twice that in Fig.~\ref{cl1018}.  The aperture densitometry significance
  contours of panels D and H use a $1\farcm 5$ aperture radius.}
\label{cl1054-12}
\end{figure*}

\begin{figure*}
\sidecaption
\includegraphics[width=12cm]{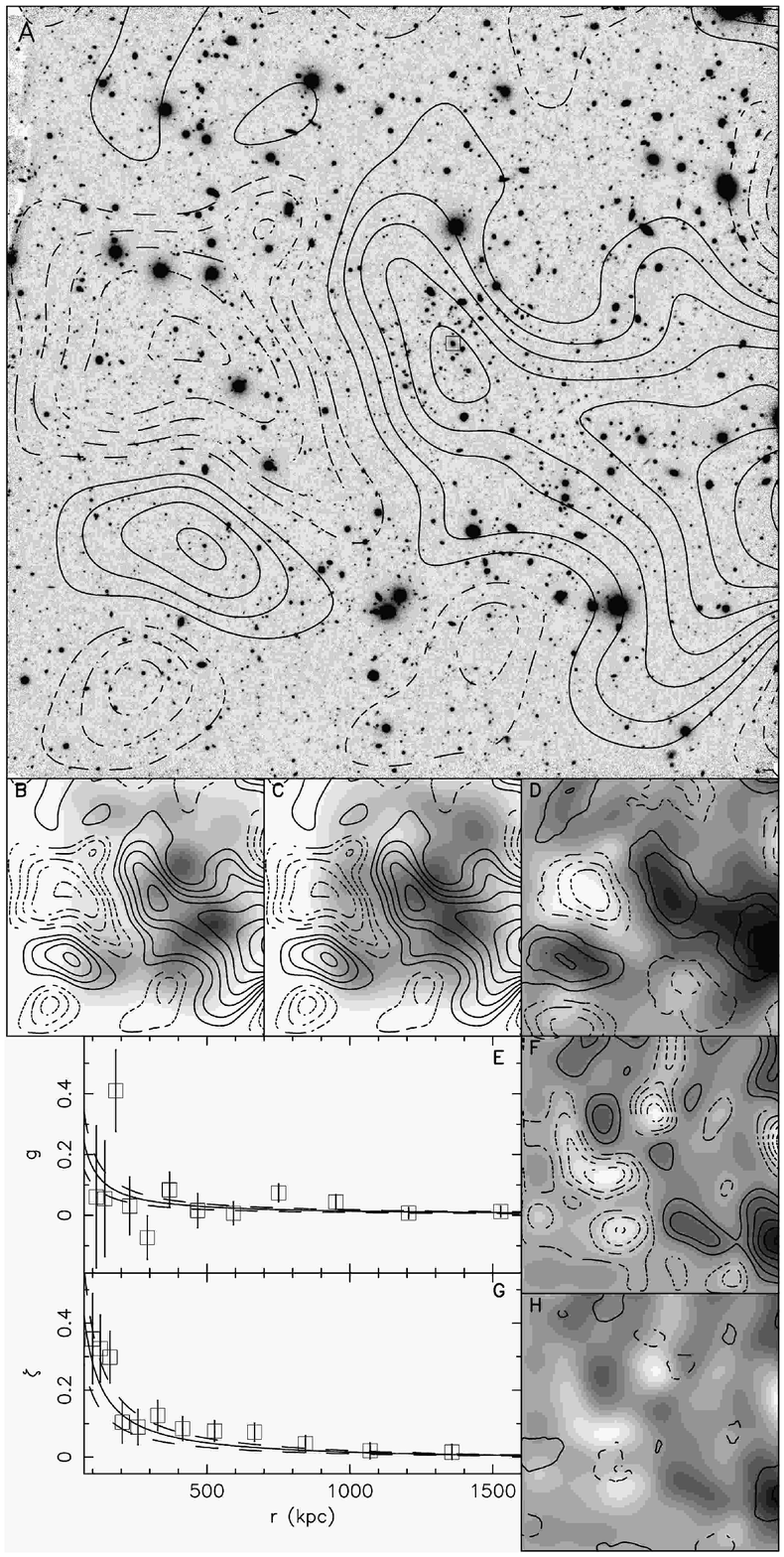}
\caption{Shown in panels A--F are the figure for CLJ1103.7$-$1245, with images
  $7\farcm 15$ or 3.40 Mpc in the cluster restframe on a side, using the same
  layout as Fig.~\ref{cl1018}.  The aperture densitometry significance
  contours of panels D and H use a $2\farcm 5$ aperture radius.}
\label{cl1103}
\end{figure*}

\begin{figure*}
\sidecaption
\includegraphics[width=12cm]{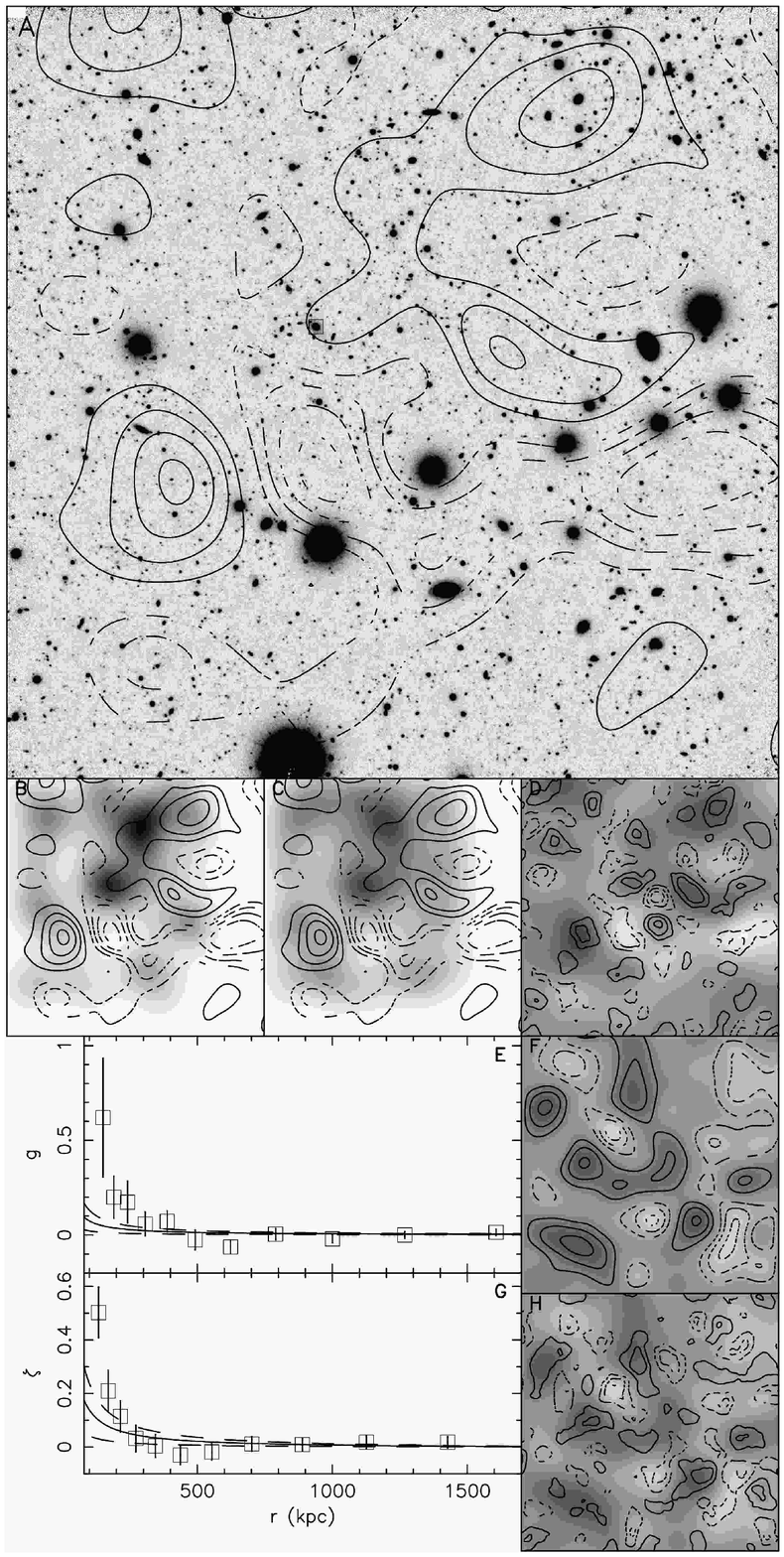}
\caption{Shown in panels A--F are the figure for CLJ1122.9$-$1136, with images
  $7\farcm 09$ or 3.21 Mpc in the cluster restframe on a side, using the same
  layout as Fig.~\ref{cl1018}.  The aperture densitometry significance
  contours of panels D and H use a $1\farcm 0$ aperture radius.}
\label{cl1122}
\end{figure*}

\begin{figure*}
\sidecaption
\includegraphics[width=12cm]{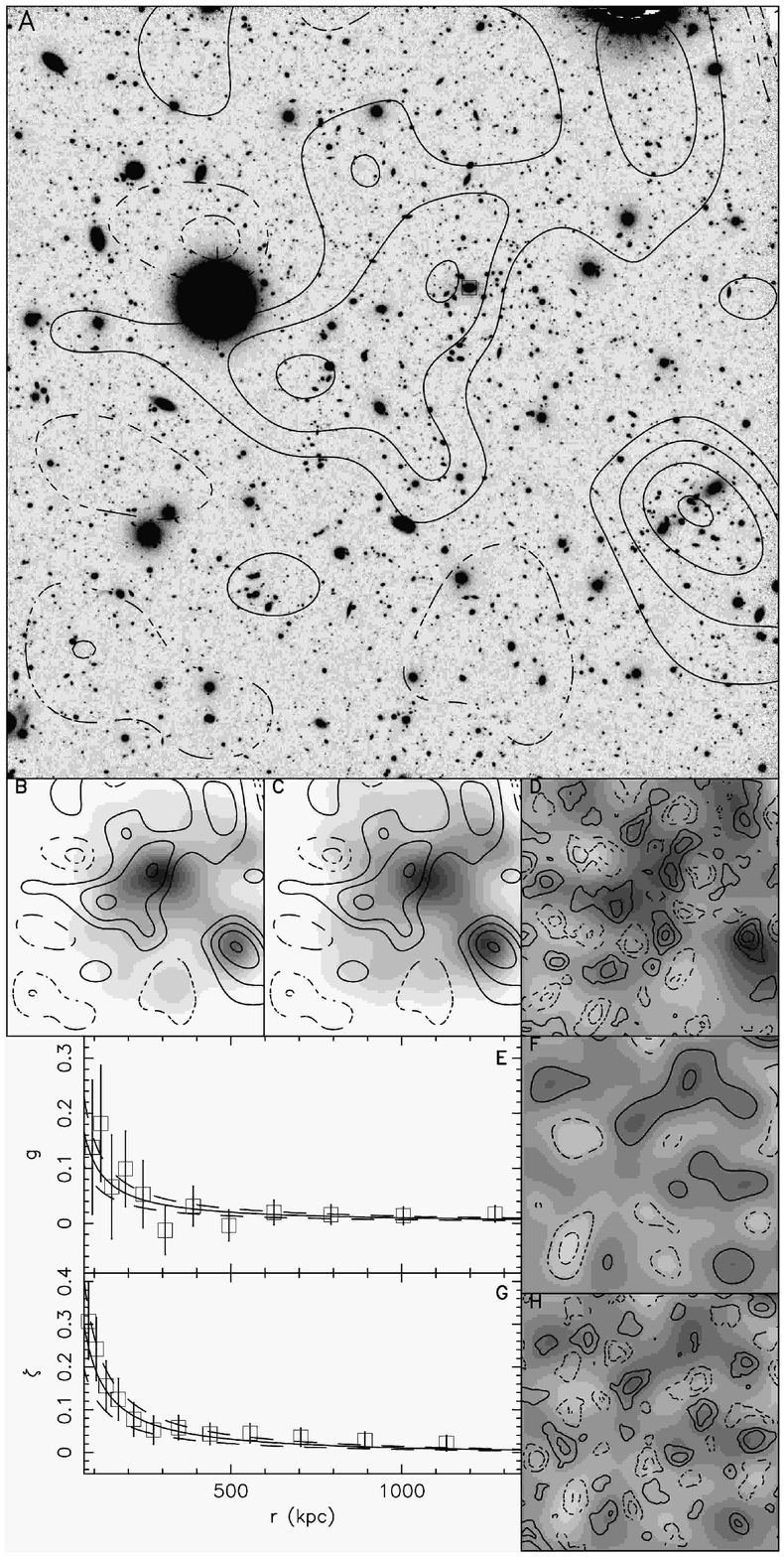}
\caption{Shown in panels A--F are the figure for CLJ1138.2$-$1133, with images
  $7\farcm 04$ or 2.52 Mpc in the cluster restframe on a side, using the same
  layout as Fig.~\ref{cl1018}.  The aperture densitometry significance
  contours of panels D and H use a $1\farcm 0$ aperture radius.}
\label{cl1138}
\end{figure*}

\begin{figure*}
\sidecaption
\includegraphics[width=12cm]{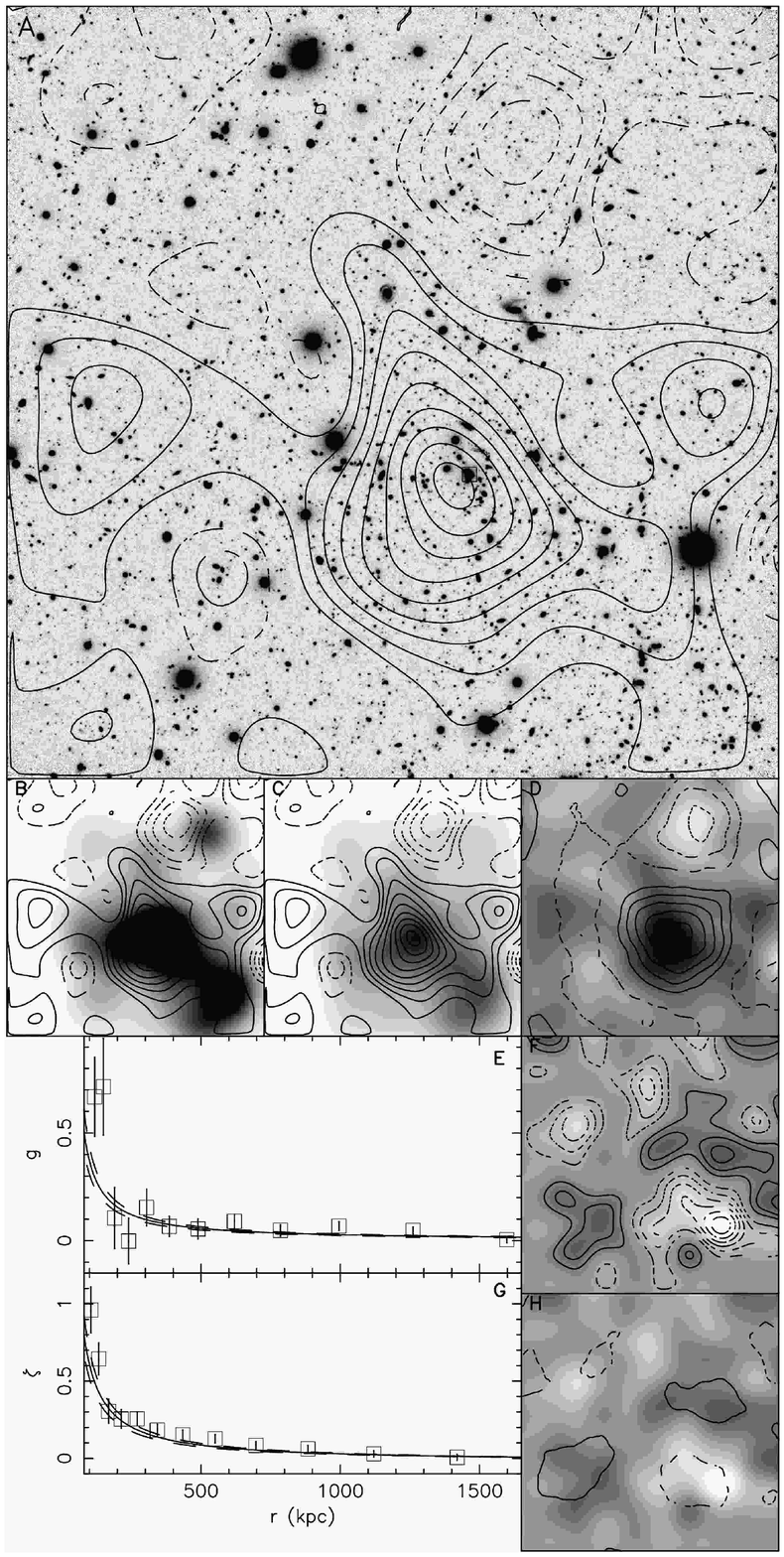}
\caption{Shown in panels A--F are the figure for CLJ1216.8$-$1201, with images
  $7\farcm 09$ or 3.19 Mpc in the cluster restframe on a side, using the same
  layout as Fig.~\ref{cl1018}.  The greyscale in panels B and C have their
  maximum value at twice that in Fig.~\ref{cl1018}.  The aperture densitometry significance
  contours of panels D and H use a $3\farcm 0$ aperture radius.}
\label{cl1216}
\end{figure*}

\begin{figure*}
\sidecaption
\includegraphics[width=12cm]{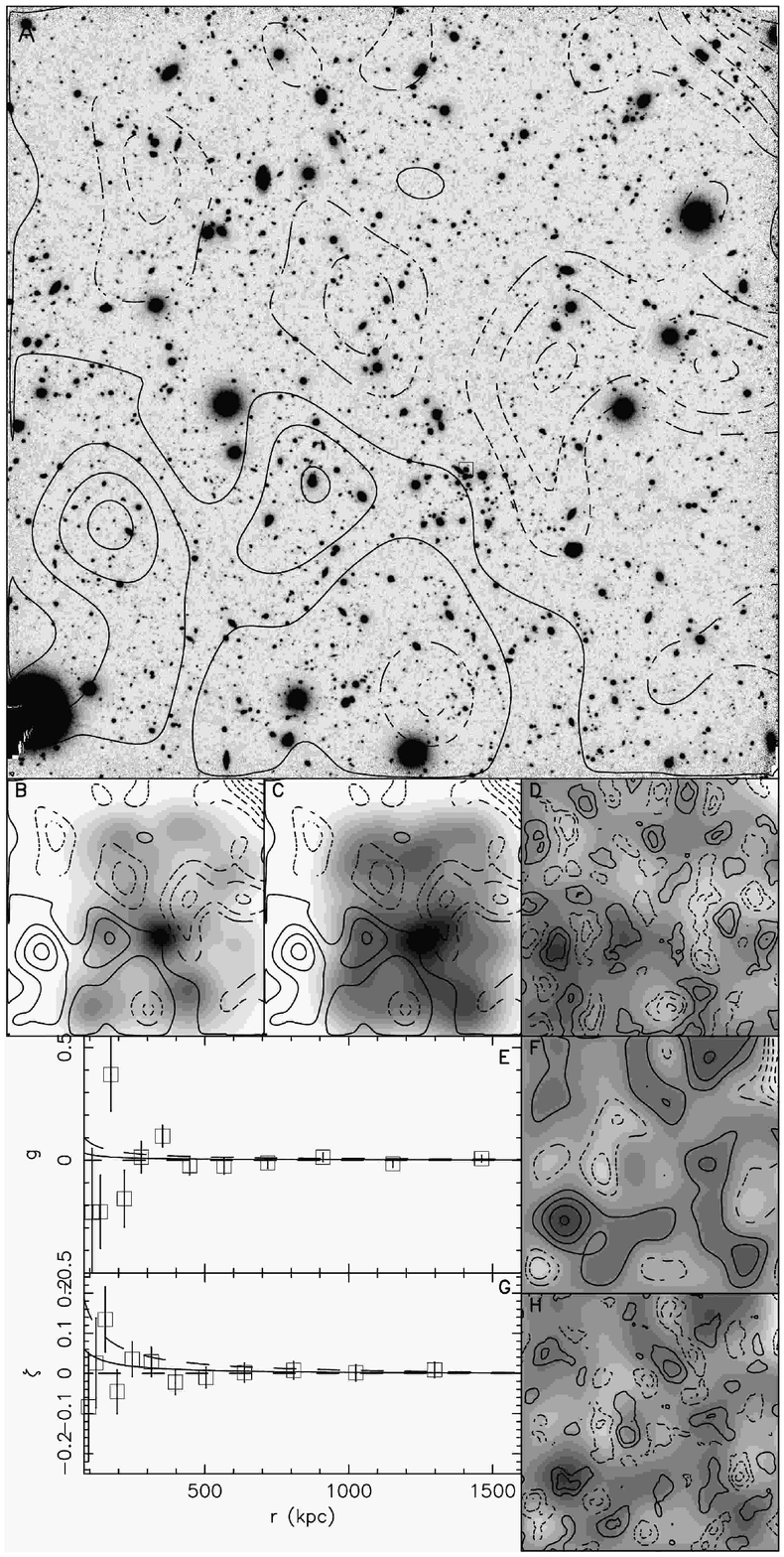}
\caption{Shown in panels A--F are the figure for CLJ1227.9$-$1238, with images
  $7\farcm 15$ or 2.94 Mpc in the cluster restframe on a side, using the same
  layout as Fig.~\ref{cl1018}.  The aperture densitometry significance
  contours of panels D and H use a $1\farcm 5$ aperture radius.}
\label{cl1227}
\end{figure*}

\begin{figure*}
\sidecaption
\includegraphics[width=12cm]{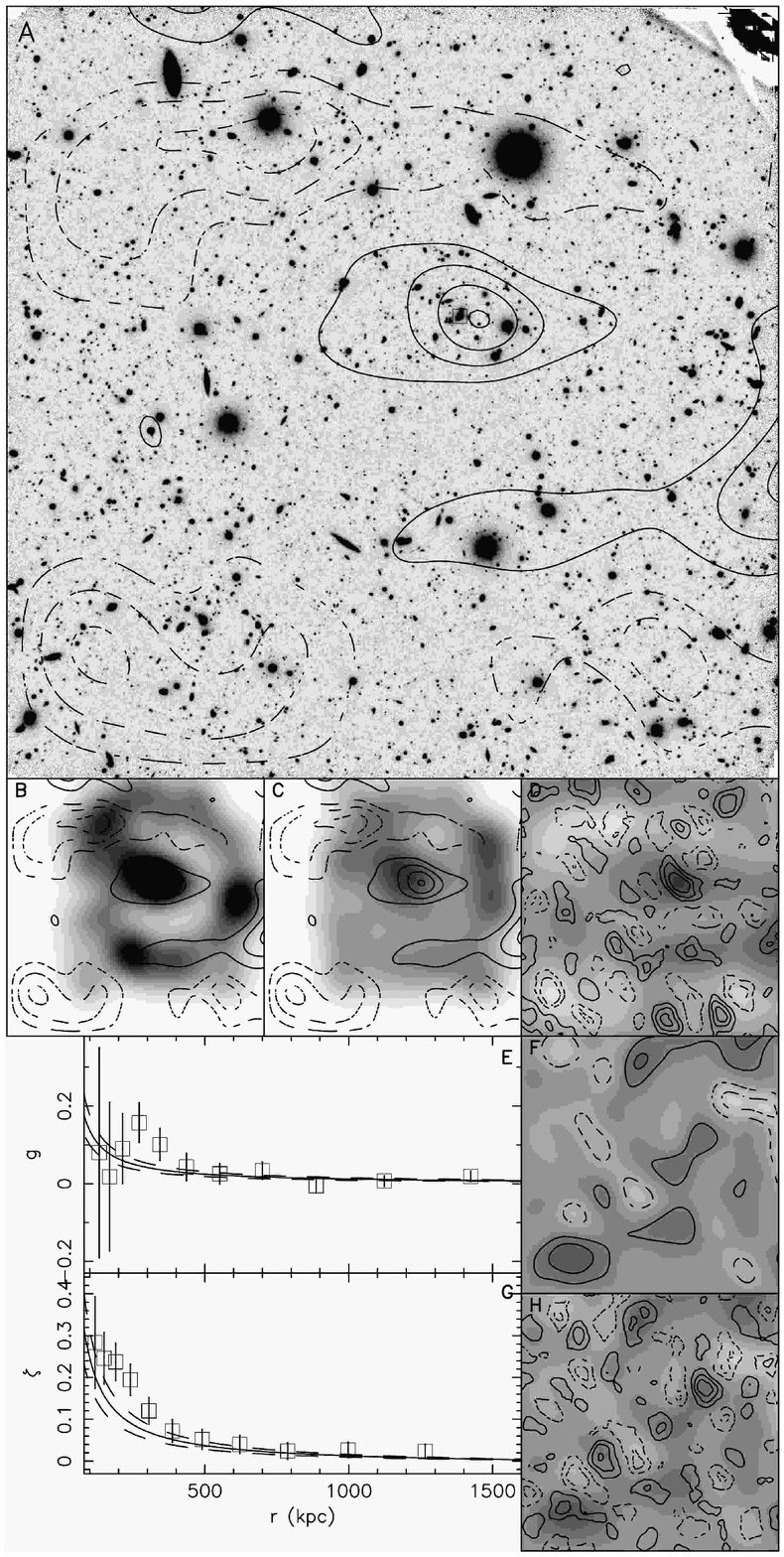}
\caption{Shown in panels A--F are the figure for CLJ1354.2$-$1230, with images
  $7\farcm 15$ or 3.16 Mpc in the cluster restframe on a side, using the same
  layout as Fig.~\ref{cl1018}.  The aperture densitometry significance
  contours of panels D and H use a $1\farcm 0$ aperture radius.}
\label{cl1354}
\end{figure*}

The weak lensing results for the high-redshift sample are shown in Figs~12--21, using the
same layout as the intermediate redshift sample.  The greyscale range in panels B and C are
the same as for the intermediate redshift sample except for CLJ1054.4$-$1146, CLJ1054.7$-$1245,
and CLJ1216.8$-$1201, which have the maximum in the scale range at twice the normal value.
A summary of the weak lensing data for each cluster can be found in Table \ref{hztable}.
A description of each cluster's lensing analysis follows.

\subsection{CLJ1037.9$-$1243}

In this field are two clusters, a $z = 0.580$ cluster located slightly north-east of the
center of the field and a $z = 0.43$ cluster located in the southwest portion of the
field.  Both clusters are detected in the weak lensing reconstruction at moderate
significance.  The north-eastern cluster, which is the one identified in the LCDCS as
being at high-redshift, is best fit with a $\sigma_\mathrm{v} = 503^{+84}_{-102}$ km/s SIS
model and is detected at a $2.7\sigma$ significance in the shear profile and $3.0\sigma$
with $M_{\mathrm{ap}}$.  The south-western cluster is best fit with a 
$\sigma_\mathrm{v} = 513^{+61}_{-72}$ km/s SIS
model and is detected at a $3.7\sigma$ significance in the shear profile and $2.8\sigma$
with $M_{\mathrm{ap}}$.  Both of these mass measurements are presumably biased toward
higher masses by the other peaks.  Attempts at fitting both peaks simultaneously results
in a large covariance between the two peaks, due mainly to the small field size not
providing a good constraint on the total mass.  Additionally, in the lensing reconstruction
is a northern extension from the north-eastern cluster which continues until the edge of the
image.  This northern extension is also found in the cluster galaxy distribution.
The sharp rise in the surface mass density in the reconstruction at the northern edge of
this extension is not found in
the mass aperture significance map, and is likely due to the increased noise level at the
edges of the reconstruction.

\subsection{CLJ1040.7$-$1155}

No significant peaks are detected in this field.  Using the BCG as the center of the cluster
we find a best-fit mass model which has positive mass, but is significant only at
$1.3\sigma$ in shear and $1.0\sigma$ with $M_{\mathrm{ap}}$.  The $2\sigma$ upper limit
on the velocity dispersion for this cluster is 684 km/s.  The weak lensing mass measurement
is in good agreement with the $\sigma_\mathrm{v} = 418^{+55}_{-46}$ km/s velocity dispersion
measured from spectra of 30 cluster members \citep{HA04.1}.

\subsection{CLJ1054.4$-$1146}

A $\sigma_\mathrm{v} = 885^{+99}_{-117}$ km/s mass peak is detected near the BCG 
of CLJ1054$-$1146 ($z = 0.697$) at $3.8\sigma$ significance in shear and $2.9\sigma$ 
with $M_{\mathrm{ap}}$.  Additional peaks are
found $3\arcmin$ south-southwest of the cluster, $3\arcmin$ south-southeast of the cluster, 
$1\farcm5$ east-southeast of the cluster, and $1\arcmin$ northeast of the cluster.
The SSW cluster is detected at the $2.4\sigma$ level with $M_{\mathrm{ap}}$, the other three
at $\sim 2\sigma$.  The SSW peak is located near a galaxy
group with photometric redshifts of 0.25, while the two
southeastern peaks are located near large, unconcentrated galaxy overdensities
with photometric redshifts consistent with cluster galaxies.  The northwestern
peak does not have any galaxy overdensities nearby.  In addition, there is a long
filamentary like extension starting $\sim 1\arcmin$ south of the cluster and running
due west, which overlaps two galaxy overdensities, one of which is likely to be around $0.8<z<1.0$
while the other is at lower redshift than the cluster, from the photometric redshifts.
The weak lensing
mass measurement is much larger than the $\sigma_\mathrm{v} = 589^{+78}_{-70}$ km/s
velocity dispersion measured from spectra of 48 cluster members \citep{HA04.1}.  This mass
discrepancy, combined with the large swath of similar redshift galaxies to the east 
of the cluster and that the mass-to-light ratio from the weak lensing mass measurements
are consistent with the sample average, suggests that a large amount of mass is currently 
infalling into
the cluster.  Such mass would be included in the weak lensing estimates but would
not have yet affected the velocity dispersion of the cluster galaxies.  A large,
diffuse screen of mass would also explain why the $M_{\mathrm{ap}}$ significance
is lower than that from the shear profile and why the highest $M_{\mathrm{ap}}$
significance is from the smallest filter radius used.

\subsection{CLJ1054.7$-$1245}

CLJ1054$-$1245 ($z = 0.750$)  is detected as the fourth most massive cluster in
the EDisCS sample, with a $\sigma_\mathrm{v} = 906^{+88}_{-102}$ km/s velocity
dispersion at $4.4\sigma$ significance in shear and $4.9\sigma$ with $M_{\mathrm{ap}}$.
Three additional, less significant peaks are located $\sim 2\arcmin$ west, north
and northeast of the cluster.  The northeastern peak is located near an overdensity of
galaxies with photometric redshifts of $\sim 0.6-0.66$.  
The western peak is located near an overdensity of lower redshift galaxies, while
the northern peak is not
obviously associated with any large galaxy grouping.  The weak lensing cluster
mass is much larger than the spectroscopic velocity dispersion $\sigma_\mathrm{v}
= 504^{+113}_{-65}$ km/s \citep{HA04.1}.  
 One possible explanation for the discrepancy would
be that the cluster has a triaxial mass distribution with a dominant major axis,
and that this axis is closely aligned to the line-of-sight through the cluster.  This
would result in a large amount of mass being projected into a small area centered on
the cluster BCG, and produce an overestimate of the central density of the cluster in 
the weak lensing mass measurements.  Such alignments in simulated clusters result in
large overestimates of the dynamical mass of the cluster \citep{CL04.1}.
Another possible explanation stems from the spectroscopic survey, which shows that
this cluster may have an extended tail toward high-redshift, and that many of
these galaxies in this high-redshift tail are spatially coincident in
projection with the cluster core.  This may indicate an extended filamentary
structure along the line-of-sight associated with the cluster, with the weak
lensing observations including all of the mass of the filament. { The mass-to-light
ratio of the cluster, however, is consistent with the sample average, suggesting
that the high weak lensing mass is not an overestimate due to noise.
Using the velocity dispersion of the cluster galaxies to calculate the mass, results 
in a mass-to-light ratio which is significantly lower than the sample average.
There are also two probable strong lensing giant arcs in the cluster, both located
north of the BCG.  The brighter of these two is at a distance of $17\arcsec$ from
the BCG, which is consistent with the weak lensing mass measurement if this galaxy
is at $z\sim 1.5$.  We do not have a spectroscopic redshift for this arc, but
photometric redshifts give it a best-fit redshift of 1.7, with an allowable range
of 1.2 to 1.8.}

\subsection{CLJ1103.7$-$1245}

{ At $z=0.96$, CLJ1103.7$-$1245 is the highest redshift cluster in the EDisCS sample.
Using the chosen BCG as the mass centroid results in a moderately significant mass peak 
($3.0\sigma$ in shear and $2.9\sigma$ with $M_{\mathrm{ap}}$).  Spectroscopy has shown,
however, that there are two additional galaxy concentrations near this cluster, one
to the southwest of the cluster at $z=0.63$ and one to the northwest at $z=0.70$.
The $z=0.70$ structure is not detected in the weak lensing mass reconstruction, while
the $z=0.63$ structure is likely responsible for the large $\kappa$ peak at the western
edge of the mass reconstruction.  Even though the $z=0.63$ structure has a much larger
$\kappa$ value in the reconstruction, due to the difference in $\Sigma_{\mathrm{crit}}$
the $z=0.96$ cluster is the most massive structure in the reconstruction.  The measured
mass for the cluster, however, is likely overestimated due to the inclusion of the surface
densities of the other two structures in the radial shear measurements, as can be seen
by the bump in $\kappa$ from the aperture densitometry measurements in panel G of
Fig.~\ref{cl1103} between 400 and 700 kpc (which is the separation between the cluster
and the $z=0.63$ structure).  The extremely high mass-to-light ratio measured for this
cluster is likely due in part to the overestimation of the mass due to the projected
structures and an underestimation of the total light, as compared to the lower redshift
clusters, due to the incompleteness limit in the luminosity profile occurring at a
brighter absolute magnitude.  It should also be noted that the $z=0.96$ cluster is
also being lensed by the $z=0.63$ structure, which will result in it being displaced
by a few arcseconds from its true position and a magnification of the cluster galaxies by
$\sim 0.1$ magnitudes.}

\subsection{CLJ1122.9$-$1136}

No significant peaks are detected in this field.  Using the brightest
galaxy in a small group near where the LCDCS peak is located gives a
best-fit $\sigma_\mathrm{v} = 589^{+172}_{-262}$ km/s SIS profile
at $1.4\sigma$ significance in shear and $2.3\sigma$ with $M_{\mathrm{ap}}$.

\subsection{CLJ1138.2$-$1133}

A moderately significant mass peak is located near the BCG of CLJ1138$-$1133
($z = 0.480$) with $\sigma_\mathrm{v} = 529^{+78}_{-96}$ km/s velocity
dispersion at $2.9\sigma$ significance in shear and $1.4\sigma$ with
$M_{\mathrm{ap}}$.  Two additional, smaller peaks are located $1\farcm 5$
northeast and $2\arcmin$ southeast of the cluster, both near galaxy
overdensities of similar redshift to the cluster.  Another more
significant peak is located $4\arcmin$ southwest of the cluster and is
spatially coincident with an overdensity of lower redshift galaxies.
The two peaks near the cluster are likely the result of the small
filter radius for the highest significance $M_{\mathrm{ap}}$ measurement
and its lower significance compared to that of the shear profile.  The
mass of these two peaks are included in the total mass of the system
at radii larger than their separation from the cluster core.  

\subsection{CLJ1216.8$-$1201}

CLJ1216-1201 ($z = 0.794$) is the most massive cluster in the EDisCS sample,
with a best-fit SIS model with $\sigma_\mathrm{v} = 1152^{+70}_{-78}$ km/s
and a $6.8\sigma$ significance in the shear profile and $5.2\sigma$
with $M_{\mathrm{ap}}$.  A three-image, strong lensing arc system is also 
located in the cluster at a $\sim 15\arcsec$ radius from the BCG.
The weak lensing mass is in reasonable agreement with the $\sigma_\mathrm{v}
= 1018^{73}_{-77}$ km/s velocity dispersion measured from spectra of 67 cluster
members \citep{HA04.1}.  The cluster galaxies have a northern and south-western 
filamentary-like extension, both of which are detected at moderate significance 
in the mass reconstruction.  The mass from these filaments would be located in
the negative weight portion of the $M_{\mathrm{ap}}$ filter, which would explain
the lower significance compared to the shear profile.

\subsection{CLJ1227.9$-$1138}

A marginally significant mass peak is located in this field, with $2.3\sigma$
significance in shear and $1.8\sigma$ significance with $M_{\mathrm{ap}}$,
but is $1\farcm 5$ displaced from the BCG of CLJ1227.9$-$1138 ($z = 0.634$).  The
peak does not appear to be associated with any overdensity of galaxies (the
bright, extended object near the center of the peak in Fig~\ref{cl1227} is
is actually two blended stars).  Using the BCG as the center of a shear
profile results in a best-fit SIS model with $\sigma_\mathrm{v} = 
308^{+221}_{-308}$ km/s at a $0.5\sigma$ significance in the shear and
$1.2\sigma$ significance with $M_{\mathrm{ap}}$.  The $2\sigma$ upper limit
on the velocity dispersion is 600 km/s.  

\subsection{CLJ1354.2$-$1230}

A moderately significant mass peak is located near the BCG of CLJ1354.2$-$1230
($z = 0.598$) with a $\sigma_\mathrm{v} = 747^{+87}_{-98}$ km/s velocity
dispersion at $3.9\sigma$ significance in shear and $3.7\sigma$ with
$M_{\mathrm{ap}}$.  The broad, filamentary-like structure seen in the
mass reconstruction south of the cluster breaks up into several $2\sigma$
peaks in $M_{\mathrm{ap}}$ measurements, and is located near a small group of
lower redshift galaxies.  

\section{Discussion}

\begin{figure*}
\sidecaption
\includegraphics[width=12cm]{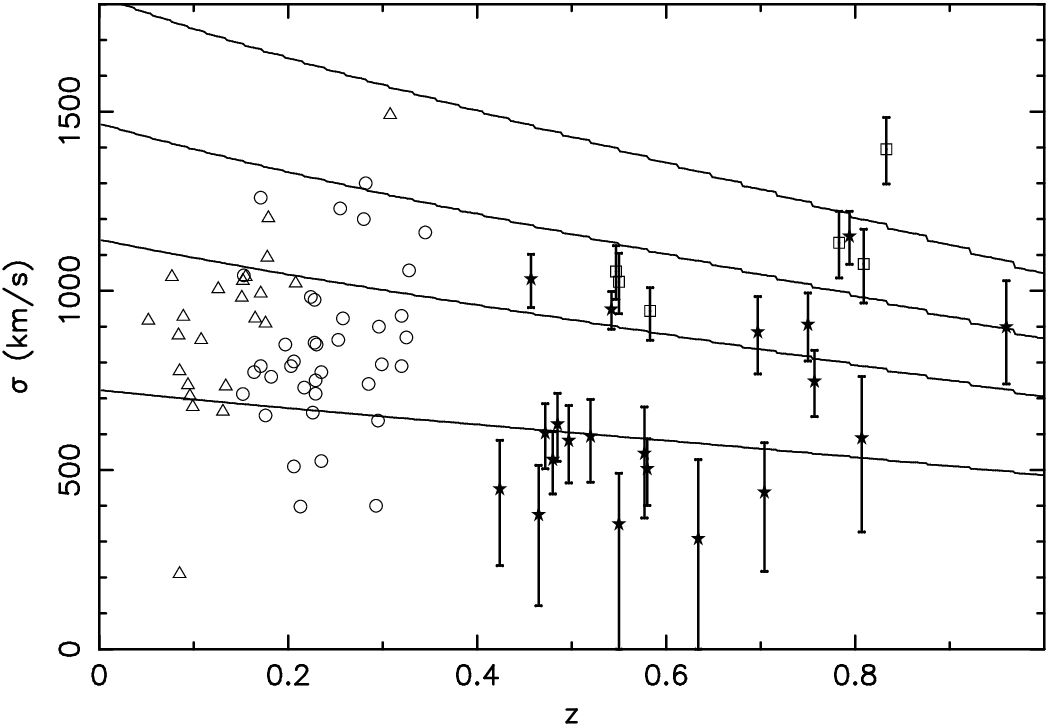}
\caption{
Plotted above are the best-fit SIS velocity dispersions for the EDisCS clusters (stars),
high-$z$ X-ray selected clusters from \citet{CL00.1} (squares), and lower redshift 
X-ray selected cluster samples from \citet{DA02.1} (circles) and \citet{CY04.1} (triangles).
The error bars for the EDisCS and high-$z$ X-ray selected samples are given as $1\sigma$,
but the measured quantity is proportional to the square of the velocity dispersion.
The lines show a model for the evolution of the cluster mass with redshift from
\citet{WE02.1}.
}
\label{velplot}
\end{figure*}

Because the mass reconstructions are the combination of the true mass surface
density field and a noise field, the position and shape of the cluster
mass peak can be greatly altered by the noise field.  As such, if one were
to pick the centroid of the mass peak in the 2-D reconstruction as the center
of mass of the cluster, the resulting best-fit mass profile would tend to
overestimate the cluster mass as one is picking the highest noise peak 
which is near the true mass peak.  We have instead assumed that the BCG is
the center of cluster for making the mass and significance measurements
in Tables \ref{hztable} and \ref{lztable}.  For relaxed clusters, N-body
simulations suggest that the BCG is a good choice for the center of the
cluster \citep{FR96.1}.  Because these clusters are located
in the epoch of massive cluster formation \citep[eg.][]{EK96.1}, however, there is a strong likelihood
that many of the clusters are undergoing a major merger event and that
the BCG may not be located at the center of mass of the cluster.  For such
cases we will be underestimating the true mass and significance of the
cluster.

While only eight of the twenty clusters are detected at greater than $3\sigma$
significance, all of the clusters are detected with a best-fitting mass model
which has positive mass (positive $\sigma ^2$ in the SIS fits).  
If there was not a large mass concentration associated
with the clusters then half of the clusters, on average, would be detected
with best-fitting mass models which have negative mass.  Thus, while we cannot
conclude with any confidence that the seven clusters which we detect at less
than $2\sigma$ significance (CLJ1119.3$-$1129, CLJ1202.7$-$1224, CLJ1238.5$-$1144,
CLJ1353.0$-$1137, CLJ1040.7$-$1155, CLJ1122.9$-$1136, and CLJ1227.9$-$1138)
individually have an associated mass overdensity, collectively we measure a
$3\sigma$ significance that, on average, these clusters have mass associated with
them.  This significance was calculated both from the probability that a
Gaussian error function on a zero mass model would produce seven positive mass
measurements at or greater than the measured significances, and from stacking
the background galaxy catalogs of the seven clusters after aligning to a 
common BCG position and measuring the significance of the resulting best-fit
mass model.

We show in Figure~\ref{velplot} the best-fit SIS velocity dispersions for the EDisCS
clusters, along with the high-$z$, X-ray selected clusters of \citet{CL00.1} and lower
redshift samples of \citet{DA02.1} and \citet{CY04.1}.  Also shown are models for the 
mass growth rate of clusters from \citet{WE02.1}.  In these models, the clusters
grow as $\propto \exp(-\alpha z)$, with massive clusters having typical $\alpha$ values
measured in simulations between 1.1 and 1.6.  The plotted lines assume a value of
$\alpha = 1.4$, which indicates a formation redshift of $z\sim 1.05$
\citep{BO02.1}.  The change in the measured velocity dispersion was calculated
by converting the cluster mass into a characteristic radius and concentration via
the equations in \citet{NA97.6}.  These were then used to calculate the reduced shear
profile for the cluster, which was then fit over the same radial range as the
measured data with a projected SIS model. The two lower redshift samples
were selected to have high X-ray luminosity, and are representative of the most massive
clusters in the nearby, $0.1\ga z \la 0.3$, universe.  The total sky area covered
by the X-ray surveys which found the high-$z$ clusters shown in Figure~\ref{velplot}
is $\sim 200 \Box ^\circ$ \citep{HE92.1,HE01.1}, or roughly 1.5 times that of the LCDCS.
As a result, we would expect to find in EDisCS on order of 2 clusters of mass comparable
to the X-ray selected clusters in each of the two redshift groupings of the sample.  
{ There are two clusters in the $z\sim 0.5$ sample and one in the $z>0.6$ sample
which have measured masses comparable with the X-ray selected clusters, in statistical
agreement with expectations.}

As can be seen in Figure~\ref{velplot}, the X-ray selected, high-$z$ clusters are only comparable with the
highest mass tip of the lower redshift cluster mass distribution.  The EDisCS clusters are more
comparable in mass range to the lower redshift distribution, but do have a bias
toward finding higher mass clusters at higher redshift.  The mass comparison with
the lower redshift clusters, however, do come with a number of caveats.  First,
the \citet{WE02.1} models were only measured for clusters with $z=0$ masses of up
to $10^{15} M_\odot$, while the $z=0$ masses for many of the EDisCS clusters from these
models will wind up being several times that.  Second, the plotted velocity
dispersions are measured by fitting the azimuthally-averaged shear profiles around
the cluster centers with a projected spherical SIS shear profile, but as the area
on the sky observed around both sets of clusters was of similar size, the fit for the
EDisCS clusters goes out to much larger radius than that of the lower redshift sample.
As SIS models are known not to provide a good fit to clusters at large radii
\citep{CL01.1,CL02.1}, the differing scales for the fits may provide a bias which
is dependent on the redshift of the cluster.  

The significance of the secondary peaks in the mass reconstructions is best determined 
with the $M_{\mathrm{ap}}$ statistic, { which, because it is a compensated filter,
is relatively insensitive to the presence of the cluster in the field provided the cluster
mass centroid is not within the aperture}.  The mass aperture statistic has been shown to
have Gaussian errors regardless of the ellipticity distribution of the background
galaxies \citep{SC96.3}, and has the property that moving the center of the aperture by half
of a filter radius results in a mass measurement which is uncorrelated with
the previous peak.  { Like the mass reconstructions, however, the mass aperture statistic
is a combination of the true signal and a noise field, so simply looking for peaks in
an oversampled map, such as that shown in panels d and h in Figs.~2--21, will result in
peaks being measured as more significant they they truly are on average.  Instead, to obtain
an unbiased measurement of the structure's $M_{\mathrm{ap}}$ value and significance, one
must obtain an independent measurement of the position of the center of mass for the
structure.  One can, however, make a statistical measurement of the amount of structure
in a given field using a grid of $M_{\mathrm{ap}}$ measurements, with each measurement
point separated from the others by at least half the filter radius.}

{ We constructed $M_{\mathrm{ap}}$ grids using the filter radii given in
Tables \ref{lztable} and \ref{hztable} for each cluster, positioning the grid such
that one point was placed on the centroid of the BCG.  We then excluded that point
and the 12 nearest points (those which had the BCG within the radius of the 
$M_{\mathrm{ap}}$ filter) from the measurements and counted the number of grid
points which had $M_{\mathrm{ap}}$ significances of greater than $2\sigma$.}
For a $7\arcmin \times 7\arcmin$ field one would expect
to find, on average, 0.69, 1.0, 1.56, 2.77, and 6.23 positive $M_{\mathrm{ap}}$ values with 
greater than $2\sigma$ significance using $3\farcm 0, 2\farcm 5, 2\farcm 0, 
1\farcm 5,$ and $1\farcm 0$ filter radii respectively.  On average, we find each field
has 1.7 times the number of $2\sigma$ or greater significance values than expected from
the noise in the $M_{\mathrm{ap}}$ maps, suggesting that $\sim 60\%$ of the points
are due to noise and the other $\sim 40\%$ are real mass structures along the line of sight.
There are four times as many $3\sigma$ or greater significance points which are not
spatially coincident with the cluster as would be expected from noise, suggesting
that $\sim 75\%$ of these are real mass overdensities.  Only one of these $3\sigma$
mass points, in CL1420.3$-$1236, is not spatially coincident with a galaxy overdensity.
{ It should be noted, however, that this method will overestimate the number of detected
points at a given significance level as compared to a noise-free model of a given power
spectrum due to more low-mass peaks being scattered to higher significance than
high mass peaks being scattered to lower significance.  As such, any comparison of
such counts with models will need to accurately model the noise properties of the data.}

The weighted mean mass-to-light ratio over the sample is $133\pm 35 
\mathrm{M}_{\odot}/\mathrm{L}_{\odot}$ in rest-frame $I$ and $175\pm 46
\mathrm{M}_{\odot}/\mathrm{L}_{\odot}$ in rest-frame B.  Two clusters, CL1040.7$-$1155
and CL1227.9$-$1148,
have mass-to-light ratios (in both passbands) which are more than $1\sigma$ below the mean.
Three clusters have mass-to-light ratios that are more than $1\sigma$ above the mean.
Two of these clusters (CL1059.2$-$1253 and CL1103.7$-$1245), however, 
have additional mass peaks which are
likely associated with galaxy overdensities at different redshifts from the cluster.  The
weak lensing mass for these clusters likely has some contribution from these additional
structures, while the photometric redshift selection of cluster galaxies excludes these
additional overdensities from the cluster luminosity measurement.  Five additional clusters
have secondary mass peaks from structures unlikely to be associated with the clusters which
are close enough to affect the weak lensing mass measurements, CL1018.8$-$1211,
CL1037.9$-$1243, CL1138.2$-$1133, CL1301.7$-$1139, and CL1354.2$-$1230.  
For CL1301.7$-$1139, the secondary structure is close enough in
redshift to the cluster that the photometric redshifts cannot distinguish between the
two, so both the mass and luminosity measurements are contaminated by the structure,
resulting in a mass-to-light ratio which is consistent with the sample mean.
CL1018.8$-$1211, CL1039.9$-$1245, and CL1138.2$-$1133 have mass-to-light ratios which are 
more than $0.5\sigma$ higher than the sample mean. 
When we exclude the seven clusters that are likely contaminated by projected structures,
the mean mass-to-light ratio for the sample becomes 
$121\pm 31 \mathrm{M}_{\odot}/\mathrm{L}_{\odot}$ in $I$ and $162\pm 42 
\mathrm{M}_{\odot}/\mathrm{L}_{\odot}$ in $B$.

\begin{figure}
\includegraphics{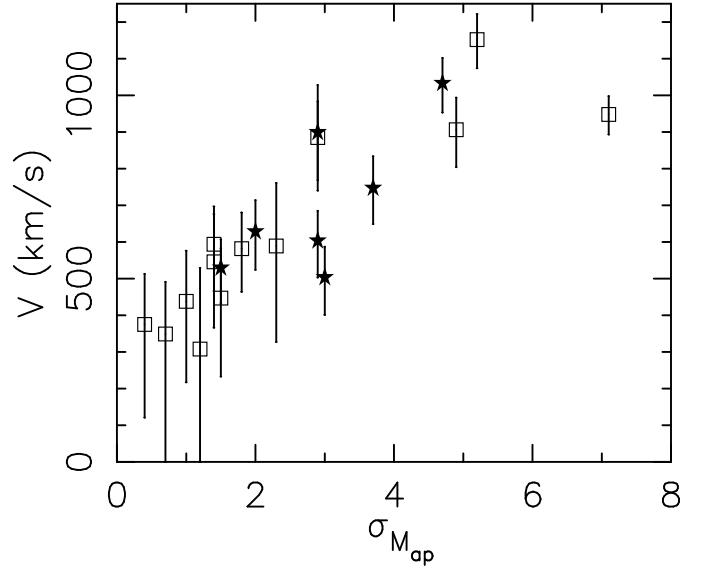}
\caption{
Plotted above are the best-fit SIS velocity dispersions for the EDisCS clusters versus
the significance of the mass aperture detection.  The solid stars show the results for
the 7 clusters with additional, unrelated mass peaks near enough to the cluster to
potentially bias the velocity dispersion measurements.  The open boxes show the
results for the other 13 clusters in the sample.
}
\label{mapvsvel}
\end{figure}

{ Fitting the mass-to-light ratio as a bi-linear function of redshift and velocity dispersion
squared gives a best fit model, taking into account the high degree of correlation in the errors
of the mass-to-light ratio and velocity dispersion squared, of
\begin{equation}
{M \over L_I} = 310^{+85}_{-90} - 280^{+120}_{-140} \ z + \left( 0.8^{+1.6}_{-2.0}\right) \times 10^{-5} \ V^2
\end{equation}
for restframe $I$ and
\begin{equation}
{M \over L_B} = 480^{+95}_{-160} - 460^{+130}_{-240} \ z + \left( 0.4^{+1.7}_{-2.4}\right) \times 10^{-5} \ V^2
\end{equation}
for restframe $B$.  The errors for each fit parameter were measured by marginalizing over the other two
fit parameters, are quoted for $68\%$ confidence limits but are not Gaussian, and there is a large 
anti-correlation between the intercept and the redshift dependent
slope which keeps the mass-to-light ratio at $z\sim 0.6$ relatively constant.
For both passbands we measure a mass-normalized brightening of the clusters at $98.4\%$ and 
$97.6\%$ significance for $I$ and $B$ bands respectively with increased redshift.  
The amount of brightening observed is broadly consistent with passive evolution models with 
starbursts at $z\sim 3-4$ for cluster galaxies
\citep[e.g.][]{BR03.1}.   We note, however, that there are several potential systematic
errors in this measurement:  First, because of the different passbands observed in the two redshift
samples, any biases in the photometric redshift selection of cluster galaxies are likely to
also be different in the two samples.  Both samples, however, have had the photometric redshifts
checked with spectroscopic redshifts and no significant difference in selection biases have
been found \citep{PE04.1}.  Second, because the mass and luminosity have been corrected
by subtraction of an outer annular measurement of a fixed size on the sky, the physical scale of
this subtracted region varies with cluster redshift.  As a result, if there is a change in the
cluster mass-to-light ratio with radial distance from the cluster center, then these corrections
could cause a bias in the measurement.  Finally, the fits are dominated by the most massive
cluster (CL1216.8$-$1201 at $z=0.794$); the removal of this cluster from the sample decreases the
slope by $\sim 15\%$ in both passbands and the significance to $90\%$ and $88\%$ for $I$
and $B$ bands.
}

{ We show in Figure~\ref{mapvsvel} the relation between the significance in the mass aperture
statistic and the best fit SIS model's velocity dispersion for each cluster.  As one would expect
due to both statistics being measured from the shear, there is a strong correlation between the
two measurements.  There is not, however, a clear distinction in the relation between the 7 clusters 
with nearby unrelated mass peaks and the other 13 clusters in the sample.  This suggests that
cluster catalogs selected using the mass aperture statistic will likely face a similar level
of contamination by projected structures as optically selected catalogs.  There is no significant
correlation between the mass-to-light ratio and the mass aperture significance, so such a catalog
would not be biased towards low mass-to-light ratio clusters, as an optically selected cluster
conceivably could be.
}

\subsection{Summary}
We present weak gravitational lensing mass reconstructions for the 20 clusters in the
ESO Distant Cluster Survey.  We show that the clusters span a large range in mass, with
the most massive clusters having a number density on the sky consistent with similar
redshift clusters found in X-ray surveys and the overall mass range similar to that
found among luminous X-ray clusters at low ($z\la 0.3$) redshifts.  
We find that 7 of the 20 clusters have
additional, unrelated structures along the line of sight which are massive enough to
be detected in the mass reconstructions and are likely to contaminate the lensing
mass measurements for the clusters.  The additional structures, however, are unlikely
to have affected the cluster detection method of the LCDCS as they are typically much
further from the cluster than the smoothing kernel used to detect the cluster light.
 We detect a positive mass for all 20 clusters,
although 7 of these are detected at less than $2\sigma$ significance.  We argue that
the lack of any best-fitting models with negative mass indicates that these poorer 
clusters still have a significant amount of mass associated with them and are
not projections of unrelated galaxies along the line of sight.

We use photometric redshifts to select likely cluster galaxies and measure the cluster
luminosity within the central 500 kpc radius.  We calculate a mass-to-light ratio for
the cluster in both restframe $I$ and $B$ passbands, and find that the clusters with
the additional structures superimposed on the cluster do, on average, have a higher
measured mass-to-light ratio.  For the remaining 13 clusters, we find that the clusters
tend to have a lower mass-to-light ratio at higher redshift, but find no change in the 
mass-to-light ratio with cluster mass.  

\begin{acknowledgements}
We wish to thank the staff of the ESO Very Large Telescope observatory in both Garching and
Paranal for their help and support in obtaining the observations.
This work was supported by the Deutsche Forschungsgemeinschaft under the project SCHN 342/3--1,
the David and Lucille Packard Foundation, and NASA under grant NAG5-13583.
\end{acknowledgements}

\bibliographystyle{aa}
\bibliography{ediscs}

\appendix
\section{Background galaxy number counts and shear noise-levels}

\begin{figure*}
\includegraphics{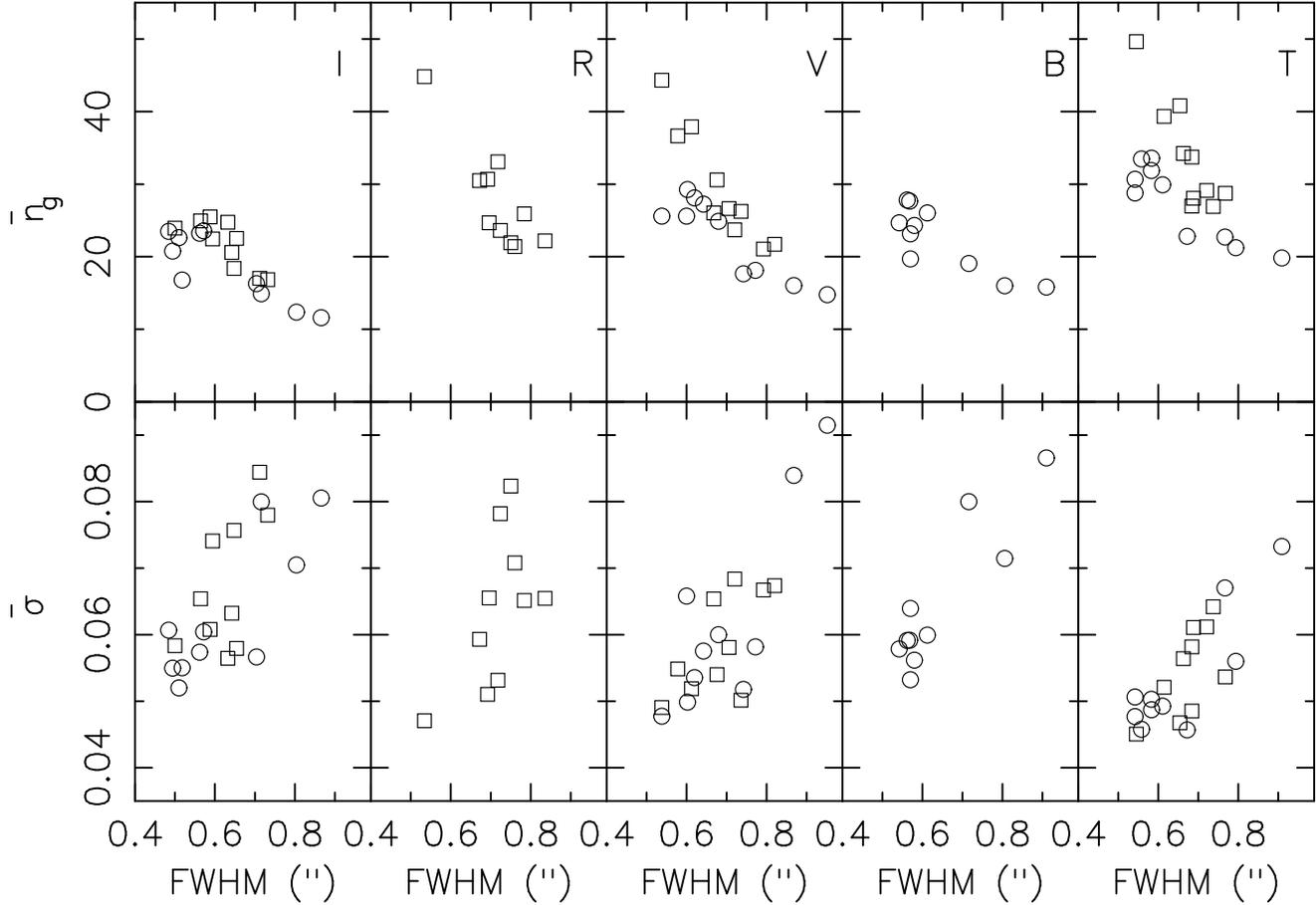}
\caption{Plotted above are the number density of background galaxies usable for shear
measurements (top panels) and the rms shear noise-level per sq.~arcminute per shear component
from these galaxies (bottom panels) as a function of PSF size for the four observed passbands
($I, R, V,$ and $B$) and the combined catalog ($T$).  The 2-hour exposure time
images of the high-$z$ clusters are plotted as squares, while the 45-minute
exposure time images of the lower redshift clusters are plotted as circles.}
\label{appfig}
\end{figure*}

Because the optical images used in this analysis were all taken with the same telescope
and instrument over a relatively short period of time using the same observing
strategy, the variation in the number density of faint, background galaxies and their 
rms shear measurements between the fields should be mostly a function of PSF size,
exposure time, bandpass, and cosmic variance without any systematic errors from varying
telescope and camera optics.  As such, this data set provides an excellent database to
study the effects that the various observing conditions can have on expected noise levels
in future ground--based surveys.

In Tables \ref{lztable} and \ref{hztable} we give the number density of background galaxies
rms shear variance in the final galaxy catalog used for the cluster mass reconstructions.
In Figure \ref{appfig} we show the number density of background galaxies and a shear-field
noise estimate (as rms shear per sq.~arcmin per shear component) as a function of PSF FWHM for each observed 
passband and the combined catalog.  In order to calculate the rms shear variation for
the galaxies, we first subtracted a smoothed shear-field, using a $10\arcsec$ Gaussian 
smoothing kernel, from each galaxies shear measurement as a first-order correction
to remove the cluster shear signal which would otherwise cause the fields around
higher-mass clusters to be detected as having a higher noise level.

As can be seen, in general the deeper 2--hour
exposures do provide a small increase in the number density of background galaxies over
the 45--minute exposures with similar PSF FWHM.  Reducing the size of the PSF, however,
results in a much larger increase in the number density of usable background galaxies,
with $0\farcs 6$ FWHM, 45-minute exposure time images having the same or more usable 
background galaxies than $0\farcs 7$ FWHM, 2-hour exposure time images.  Further,
the longer exposure time images have the greatest advantage over the shorter exposure
image images at the smallest PSF levels.
The exception is in the $I$-band, where the longer exposure images have similar
numbers of background galaxies as the shorter.  This appears to be due to three factors:
First, the initial galaxy catalogs were created from a combined image in which all of
the passbands were coadded using the inverse of the square of the sky noise level as
a weighting factor in the averaging.  As a result, the $I$-band images contributed
less weight to the final image than did the other passbands, and therefore while the
catalogs contain the highest number density of galaxies from any other image combination,
the catalogs contain fewer faint red objects than would be included in a strict $I$-band
detected catalog.  The additional galaxies in the $I$-band detected catalog, however,
are likely to be at redshifts similar to the clusters, and therefore do not contribute
much to the weak lensing signal.  The second effect is that many of the galaxies which
are picked up in the longer exposure time catalogs have measured sizes, in terms of
$r_g$, which are larger than stars in the bluer passbands but are consistent with
stars in $I$, and therefore did not have their shapes measured in $I$.  The smaller
size in $I$ than the other passbands is likely due either to the Mexican-hat filter
method for determining $r_g$ preferring smaller values at higher sky-noise levels for
a given object, or possibly to the background galaxies possessing blue outer regions.  The final effect
is that the 2-hour exposures were taken in fields with higher redshift clusters on average
than the 45--minute exposures, and as a result the color cuts to remove the cluster galaxies
in the 2--hour exposures would also have removed $z\sim 0.7-0.9$ field elliptical galaxies 
which are still present in the 45--minute exposure catalogs.

When looking at the rms shear noise-levels, however, the results seem to be a function
only of PSF FWHM and independent of the exposure time.  This is a result of the extra
galaxies detected in the longer exposure time images having a greater variance in their
shear measurements, which offsets the increased number density of the galaxies.  This
increase in the shear variance is likely to be due to three factors:  First, the fainter
objects are, on average, smaller and therefore have increased measurement noise from the
PSF correction.  Second, the fainter objects have a higher mean redshift and are thus 
observed in a bluer restframe passband, so are more likely to have their luminosity
dominated by starbursts and therefore have a higher intrinsic ellipticity.  Finally, 
because the fainter objects are at higher mean redshift, they are affected more by the
cluster's gravitational shear, but are corrected at the same level as the lower redshift
galaxies.  As a result, the higher redshift background galaxies will still have a small
increase in their rms shear variance as they are still being affected by the lensing
induced shear.  This last effect, however, should be very minor for all but the most massive
of the clusters.

It should be noted that the noise estimate in Figure \ref{appfig} is valid only for a
shear which is applied independent of the redshift of the background galaxy.  For high
redshift clusters, which cause a significantly greater shear in $z\sim 3$ galaxies than 
$z\sim 1$ galaxies, including the additional high-variance galaxies detected in the
deeper exposures does increase the signal-to-noise of the lensing measurement.

There are a number of large-scale, ground-based optical surveys which are either currently underway,
such as the CFHT Legacy Survey \citep{ME04.1}, or in planning stages, such as the LSST \citep{TY02.1}
and Pan-STARRS \citep{KA02.1}.  Almost all of these surveys have plans to measure the weak lensing shear
in the images for use in, eg, measuring the power-spectrum of mass in the nearby universe and detecting and
characterizing structures by mass.  The results given above suggest that shear signal in
these surveys will be best measured by combining the individual images of a given field which
have the lowest possible FWHM for the PSF, even if it means not using the majority of the
raw data in the image co-addition process (and presumably combining
high-seeing images together to obtain additional measurements of the shear
from the larger background galaxies).  This also implies that the surveys which plan to build
deep images by taking many shallow exposures separated by large amounts of time will be better served
by a site which delivers excellent quality seeing (FWHM $< 0\farcs 6$) for a small, but not
negligible, fraction of the time, even if it has a significant tail in the seeing distribution toward
much larger PSF size than they will by a site which delivers consistent, mediocre ($FWHM \sim\ 0\farcs8$)
image quality.

\end{document}